\documentclass[preprintnumbers,amsmath,amssymb,nofootinbib]{revtex4}
\usepackage{graphicx}% Include figure files
\usepackage{epsfig}
\usepackage{color}

\newcommand{\beq}{\begin{equation}}
\newcommand{\eeq}{\end{equation}}
\newcommand{\bea}{\begin{eqnarray}}
\newcommand{\eea}{\end{eqnarray}}
\newcommand{\bwd}{\begin{widetext}}
\newcommand{\ewd}{\end{widetext}}

\begin{document}

\title{Two-and-a-half dimensional symplectic space-charge solver 
}
%\title{A comparison of several self-consistent space-charge models for long-term tracking}

%\author{Ji Qiang\thanks{jqiang@lbl.gov}, Lawrence Berkeley National Laboratry, Berkeley, USA }

\author{Ji Qiang}
\email{jqiang@lbl.gov}
\affiliation{Lawrence Berkeley National Laboratory, Berkeley, CA 94720, USA}
%	LBNL Report Number: LBNL-2001674 
%\date{July 22, 2025}
%
\begin{abstract}
The nonlinear space-charge effect plays a significant role in high-intensity accelerators and has been extensively studied using multi-particle tracking methods. In this paper, we present a novel 2.5-dimensional symplectic space-charge solver specifically designed for long beam bunches. We begin by detailing its application to a transverse Gaussian density distribution under open boundary conditions in a straight system, where a semi-analytical expression is derived.
We then demonstrate the solver's adaptation to arbitrary distributions in open space, as well as within rectangular and round conducting pipes. Finally, we discuss the extension of this solver to circular accelerator systems. This study shows that the fast 2.5-dimensional solver can be a good 
approximation to the fully three-dimensional solver for long bunches in large circular accelerators. 

\end{abstract}

\maketitle

% Place this right after maketitle
\newcommand{\blfootnote}[1]{%
  \begingroup
  \renewcommand\thefootnote{}\footnote{#1}%
  \addtocounter{footnote}{-1}%
  \endgroup
}
\blfootnote{This version contains minor improvements over the version published in Phys. Rev. Accel. Beams (DOI: 10.1103/npx1-nclt).}

\section{Introduction}

The nonlinear space-charge effect arising from Coulomb interactions among charged particles plays a crucial role in high-intensity, high-brightness accelerators. It can lead to beam emittance growth, halo formation, and even particle losses along the accelerator. To study these space-charge effects, multi-particle tracking is employed to dynamically follow charged particles through the accelerator. In the accelerator community, most multi-particle tracking codes employ a particle-in-cell (PIC) approach to incorporate the space-charge effects self-consistently in simulations~\cite{friedman,machida,impact,track,toutatis,galambos,franchetti03,impact-t,amundson,opal}.

The self-consistent PIC method requires solving the three-dimensional Poisson equation in the beam frame at each time step, which is computationally intensive even with parallel computing resources. For a long bunch (in the beam frame) with a smooth density distribution, the variation of the electric potential along the longitudinal direction is much smaller than that along the transverse dimensions. By neglecting the longitudinal variation, the three-dimensional Poisson equation reduces to a two-dimensional one. To account for longitudinal variations in the charge density distribution, the electric potential from the two-dimensional Poisson equation is weighted longitudinally for different longitudinal slices. This approach, known as the two-and-a-half dimensional (2.5D) space-charge model in the accelerator community~\cite{synergia,pyorbit,adrian,xsuite}, does not impart a kick to the longitudinal momentum during simulation, leading to a violation of the symplectic condition in multi-particle tracking.

The gridless symplectic three-dimensional particle tracking model was previously proposed for self-consistent space-charge simulations\cite{qiang2017}. More recently, a symplectic particle-in-cell model was developed and applied to two-dimensional coasting beam simulations\cite{qiang2018}. In this paper, we propose a two-and-a-half dimensional space-charge solver that incorporates not only the longitudinal variation of the transverse momentum kick, but also the longitudinal momentum kick. By construction, this solver automatically satisfies the symplectic condition and is well-suited for long-term simulations. The longitudinal variation of the electric potential is treated as a perturbation in the three-dimensional Poisson equation, and the solution can be further refined to include this variation if necessary.

The organization of this paper is as follows: after the Introduction, we discuss the symplectic particle tracking model in Section II. Section III presents the two-and-a-half dimensional space-charge solver for a transverse Gaussian density distribution (Section III.A) and for arbitrary density distributions in a straight accelerator system (Section III.B). In Section IV, we extend the model to a circular accelerator system. Conclusions are drawn
in Section V.

\section{Symplectic Particle Tracking Model}

The charged particle dynamics inside a particle accelerator is governed by 
Hamilton's equations as:
\begin{eqnarray}
	\frac{d {\bf r}}{d s} & = & \frac{\partial H}{\partial {\bf p}} \\
	\frac{d {\bf p}}{d s} & = & -\frac{\partial H}{\partial {\bf r}} 
\end{eqnarray}
where $H({\bf r}, {\bf p};s)$ denotes the Hamiltonian of the particle using distance $s$ as an independent variable, ${\bf r}=(x,y,z=s-v_0t)$ denotes 
spatial coordinates of the particle, and ${\bf p}=(p_{x}/p_0,p_{y}/p_0,p_{z}=(p-p_0)/p_0$ the canonical momentum coordinates of the particle normalized by the reference particle momentum $p_0$ without acceleration.
Let $\zeta$ denote a 6-vector of coordinates,
the above Hamilton's equation can be rewritten as:
\begin{eqnarray}
	\frac{d \zeta}{d s} & = & -[H, \zeta] 
\end{eqnarray}
where [\ ,\ ] is the Poisson bracket. A formal solution for above equation
after a single step $\tau$ can be written as:
\begin{eqnarray}
	\zeta (\tau) & = & \exp(-\tau(:H:)) \zeta(0)
\end{eqnarray}
Here, we have defined a differential operator $:H:$ as $: H : g = [H, \ g]$, 
for arbitrary function $g$. 
For a Hamiltonian that can be written as a sum of two terms $H =  H_1 + H_2$, an approximate
solution to above formal solution can be written as~\cite{forest1}
\begin{eqnarray}
	\zeta (\tau) & = & \exp(-\tau(:H_1:+:H_2:)) \zeta(0) \nonumber \\
  & = & \exp(-\frac{1}{2}\tau :H_1:)\exp(-\tau:H_2:) \exp(-\frac{1}{2}\tau:H_1:) \zeta(0) + O(\tau^3)
  \label{ham5}
\end{eqnarray}
Let $\exp(-\frac{1}{2}\tau :H_1:)$ define a transfer map ${\mathcal M}_1$ and
$\exp(-\tau:H_2:)$ a transfer map ${\mathcal M}_2$, 
for a single step, the above splitting results in a second order numerical integrator
for the original Hamilton's equation as:
\begin{eqnarray}
	\zeta (\tau) & = & {\mathcal M}(\tau) \zeta(0) \nonumber \\
    & = & {\mathcal M}_1(\tau/2) {\mathcal M}_2(\tau) {\mathcal M}_1(\tau/2) \zeta(0)
	+ O(\tau^3)
	\label{map}
\end{eqnarray}
The above numerical integrator can be extended to $4^{th}$ order 
accuracy and arbitrary even-order accuracy following
Yoshida's approach~\cite{yoshida}.
This numerical integrator Eq.~\ref{map} will be symplectic if both the transfer map
${\mathcal M}_1$ and the transfer map ${\mathcal M}_2$ are symplectic.
A transfer map ${\mathcal M}_i$ is symplectic if and only if
the Jacobian matrix $M_i$ of the transfer
map ${\mathcal M}_i$ satisfies the following condition:
\begin{eqnarray}
M_i^T J M_i = J 
\label{symp}
\end{eqnarray}
where $J$ denotes the $6 \times 6$ matrix given by:
\begin{equation}
%\[
	J  =   \left( \begin{array}{cc}
			0 & I \\
			-I & 0
	\end{array} \right)
%\]
\end{equation}
and $I$ is the $3\times 3$ identity matrix.

For the Hamiltonian in Eq.~\ref{ham5}, we can choose $H_1$ as:
\begin{eqnarray}
	H_1 & = &  {\bf p}^2/2 + q \psi({\bf r})
\end{eqnarray}
where $\psi$ denotes potential corresponding to external fields.
This corresponds to the Hamiltonian of a group of charged particles 
inside an external field without mutual interaction among themselves. 
A single-charged particle magnetic optics method can be used to find 
the symplectic transfer map ${\mathcal M}_1$
for this Hamiltonian with the external fields from
most accelerator beam line elements~\cite{rob1}.%~\cite{rob1,alex,mad}.

We can choose $H_2$ as:
\begin{eqnarray}
	H_2 & = & K \phi({\bf r})
\end{eqnarray}
where $K = q /(p_0 v_0 \gamma_0^2)$, and $\phi$ is the space-charge potential from the solution of the Poisson's equation.
In this Hamiltonian, the effects of the direct Coulomb 
electric potential and the
longitudinal vector potential are combined together.
The electric Coulomb potential $\phi$ in the Hamiltonian 
$H_2$ can be obtained 
from the solution of the Poisson equation.
%which includes the space-charge effect and is only a function of position. 

For the space-charge Hamiltonian $H_2({\bf r})$, the single
step transfer map ${\mathcal M}_2$ can be written as:
\begin{eqnarray}
	{\bf r}(\tau) & = & {\bf r}(0) \\
	{\bf p}(\tau) & = & {\bf p}(0) - \frac{\partial H_2({\bf r})}{\partial {\bf r}} \tau
	\label{map2}
\end{eqnarray}
The Jacobi matrix of the above 
 transfer map
${\mathcal M}_2$ is 
\begin{equation}
%\[
	M_2  =   \left( \begin{array}{cc}
			I & 0 \\
			L & I
	\end{array} \right)
%\]
\end{equation}
where $L$ is a $3 \times 3$ matrix.
For $M_2$ to satisfy the symplectic condition Eq.~\ref{symp}, 
the matrix $L$ needs
to be a symmetric matrix, i.e.
\begin{equation}
	L = L^T
\end{equation}
Given the fact that $L_{ij} = \partial p_i(\tau)/\partial r_j =  - \frac{\partial^2 H_2({\bf r})}{\partial {r}_i \partial {r}_j} \tau $, the matrix $L$ will be symmetric as long as it 
is {\bf \it analytically calculated}
from the function $H_2$. 
%This is also called jolt-factorization in nonlinear
%single particle beam dynamics study~\cite{forest3}.
If both the transfer map
${\mathcal M}_1$ and the transfer map ${\mathcal M}_2$ 
are symplectic, the numerical integrator Eq.~\ref{map} for multi-particle tracking will be symplectic. 

\section{Two-and-a-half dimensional space-charge solver in a straight
accelerator}
We first discuss the solution of the Poisson equation in a straight accelerator, where the curvature is zero. This approach can also be applied to circular accelerators when the bending radius is much larger than the transverse aperture size. The effects of finite curvature will be addressed in the next section. In this case, the three-dimensional Poisson equation in the moving coordinate system can be written as:
\begin{equation}
\frac{\partial^2 \phi}{\partial x^2} +
\frac{\partial^2 \phi}{\partial y^2} + \frac{1}{\gamma^2}\frac{\partial^2 \phi}{\partial z^2}  = - \frac{\rho(x,y,z)}{\epsilon_0}
\label{poi3d}
\end{equation}
where $\gamma=1/\sqrt{1-(v_0/c)^2}$ is the relativistic factor 
and $z=s-v_0t$.
For a smooth long bunch at high energy $\frac{\partial^2 \phi}{\partial x^2} \gg \frac{1}{\gamma^2}\frac{\partial^2 \phi}{\partial z^2}$ and $\frac{\partial^2 \phi}{\partial y^2} \gg \frac{1}{\gamma^2}\frac{\partial^2 \phi}{\partial z^2}$, the longitudinal partial derivative
with respect to $z$ term can be treated as a perturbation and the Poisson equation~\ref{poi3d} can
be solved using an iterative procedure. 
%\begin{equation}
%    \phi(x,y,z) = \phi_0(x,y,z) + \phi_1(x,y,z) + \phi_2(x,y,z)+\cdots
%\end{equation}
For the zeroth-order
approximation, the above Poisson equation can be reduced into a 
two-dimensional Poisson equation:
\begin{equation}
\frac{\partial^2 \phi_0}{\partial x^2} +
\frac{\partial^2 \phi_0}{\partial y^2} = - \frac{\rho(x,y,z)}{\epsilon_0}
\label{poi2d1}
\end{equation}
The above solution can be improved in the first-order approximation
by including the partial derivative of the zeroth-order potential $\phi_0$ with respect to $z$. This results in the following two-dimension Poisson equation:
\begin{equation}
\frac{\partial^2 \phi_1}{\partial x^2} +
\frac{\partial^2 \phi_1}{\partial y^2} = - \frac{\rho(x,y,z)}{\epsilon_0} - 
\frac{1}{\gamma^2}\frac{\partial^2 \phi_0}{\partial z^2}
\label{poi2d2}
\end{equation}
This solution can be further refined by replacing $\phi_0$ with $\phi_1$ on the right-hand side of the equation and solving the two-dimensional Poisson equation again. This iterative process can be repeated multiple times until no further improvement is observed. However, for a long beam bunch with smooth longitudinal density variation, the zeroth-order approximation is often sufficient. In this work, we therefore focus on solving the zeroth-order two-dimensional Poisson equation~\ref{poi2d1}.

Now, let us consider a charge density distribution $\rho$ inside the accelerator that can be expressed as $\rho(x,y,z) = n(x,y)\lambda(z)$. The zeroth-order approximated Poisson equation can then be written as:
\begin{equation}
\frac{\partial^2 \phi_0}{\partial x^2} +
\frac{\partial^2 \phi_0}{\partial y^2} = - \frac{n(x,y)\lambda(z)}{\epsilon}
\label{poi3d1}
\end{equation}
The solution of potential from the above equation can be written as $\phi_0(x,y,z) = \varphi(x,y)\lambda(z)$, with $\varphi$
satisfying the two-dimensional Poisson equation:
\begin{equation}
\frac{\partial^2 \varphi}{\partial x^2} +
\frac{\partial^2 \varphi}{\partial y^2} = - \frac{n(x,y)}{\epsilon_0}
\label{poi2d12}
\end{equation}
If the charge density distribution function cannot be expressed as the product of transverse and longitudinal density functions, the three-dimensional density function can be approximated as a sum of longitudinal orthogonal modes (e.g., Hermite-Gaussian modes~\cite{qiang2004}):
\begin{equation}
    \rho(x,y,z) = \sum_i n_i(x,y)\lambda_i(z)
\end{equation}
The zeroth-order potential can be written as:
\begin{equation}
    \phi_0 = \sum \varphi_i \lambda_i
\end{equation}
with
\begin{equation}
\frac{\partial^2 \varphi_i}{\partial x^2} +
\frac{\partial^2 \varphi_i}{\partial y^2} = - \frac{n_i(x,y)}{\epsilon_0}
\label{poi3d1}
\end{equation}

\subsection{Space-charge solver with a transverse Gaussian density distribution}
\begin{figure}[!htb]
%\begin{figure}[htb]
%   \vspace*{-.5\baselineskip}
   \centering
   \includegraphics*[angle=0,width=230pt]{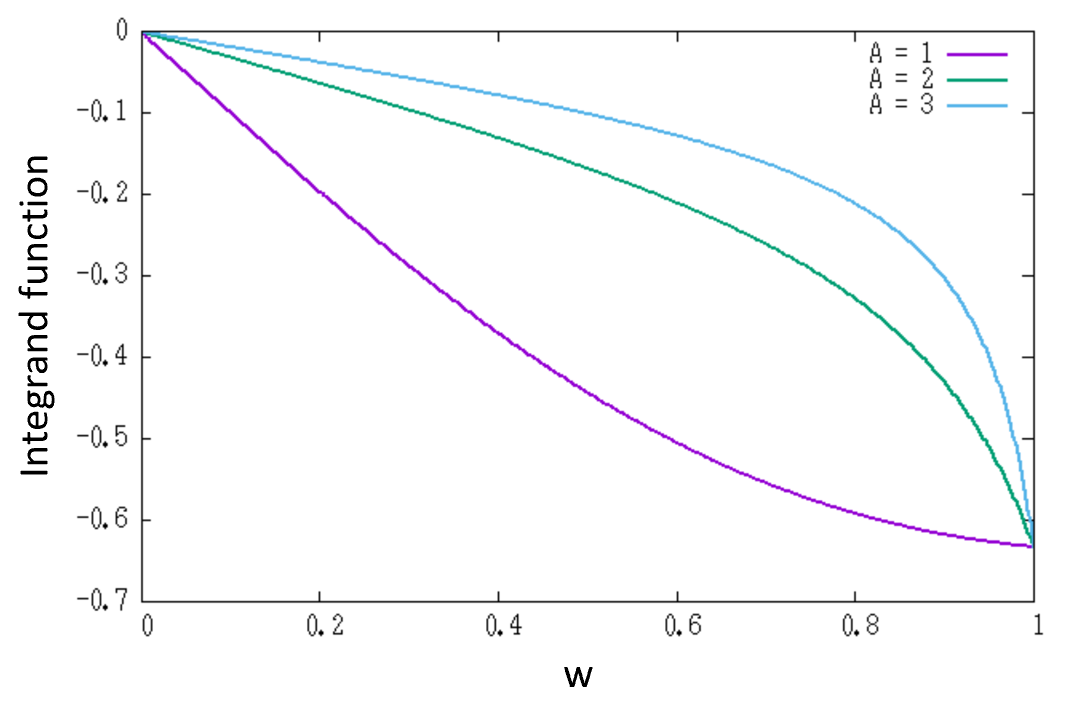}
   \caption{The integrand function in the above integral with aspect ratios $1$ (magenta), $2$ (green), and $3$ (blue) at $(x'=y'=1)$.}
   \label{integ}
%   \vspace*{-\baselineskip}
\end{figure}

%Most of beams in the high intensity accelerator show a three-dimensional Gaussian
%density distribution.
First, we consider the case in which the electric potential is subject to three-dimensional open boundary conditions. If we further assume that the transverse density distribution $n(x,y)$ is a bi-Gaussian function with standard deviations $\sigma_x$ and $\sigma_y$, respectively, the solution $\varphi$ to the above Poisson equation after regularization (removing infinite) can be written as~\cite{chao}:
\begin{equation}
    \varphi(x,y) = \frac{1}{4\pi\epsilon_0}\int_0^{\infty} dt
    \frac{\exp[-\frac{x^2}{2(\sigma_x^2+t)}-\frac{y^2}{2(\sigma_y^2+t)}]-1}{\sqrt{(\sigma_x^2+t)(\sigma_y^2+t) }    }
    \label{pint}
\end{equation}
The above integral is finite and vanishes at $(x=0,\, y=0)$. However, this potential cannot be directly used as the solution for $\phi_0$, since it yields a zero longitudinal electric field on axis ($(x=0,\, y=0)$). To satisfy the open boundary conditions, the above potential should be shifted by a constant. The resulting zeroth-order solution for the potential $\phi_0$ can be written as:
\begin{equation}
\phi_0(x,y,z) = (\varphi(x,y) + \varphi_{00})\lambda(z)
\label{phi0}
\end{equation}
where $\varphi_{00}$ is the potential at $(x=0,\, y=0)$ subject to the transverse open boundary conditions.
The integral in equation~\ref{pint} is taken over an infinite computational domain, which is not convenient for numerical evaluation. By introducing the variable substitution $w^2 = \frac{1}{1 + t/\sigma_x^2}$, the integral can be rewritten as:
\begin{equation}
    \varphi(x,y) = \frac{2A}{4\pi\epsilon_0}\int_0^{1} dw
    \frac{\exp[-\frac{1}{2}w^2x'^2-\frac{1}{2}w^2y'^2/(A^2+(1-A^2)w^2) ]-1}{w\sqrt{(A^2+(1-A^2)w^2) }    }
    \label{eq25}
\end{equation}
where the aspect ratio $A = \sigma_x / \sigma_y$, $x' = x / \sigma_x$, and $y' = y / \sigma_y$. This transformation converts the integral from an infinite domain to a finite range between $0$ and $1$. Figure~\ref{integ} shows the integrand function
in the above integral for aspect ratios $1$, $2$ and $3$. 
This integrand function varies smoothly between $0$ and $1$.
The integral can then be approximated as:
\begin{equation}
    \varphi(x,y) = \frac{2A}{4\pi\epsilon_0}(\frac{-\frac{1}{4}\delta^2(x'^2+y'^2/(A^2+(1-A^2)\delta^2))}{\sqrt{(A^2+(1-A^2)\delta^2) }}+\int_{\delta}^{1} dw
    \frac{\exp[-\frac{1}{2}w^2x'^2-\frac{1}{2}w^2y'^2/(A^2+(1-A^2)w^2) ]-1}{w\sqrt{(A^2+(1-A^2)w^2) }    })
    \label{phiapp}
\end{equation}
where $\delta$ is a small number close to zero (e.g., $0.01$). The first term on the right-hand side represents the trapezoidal rule approximation of the original integral between $0$ and $\delta$, while the second term can be efficiently computed using a quadrature rule such as Simpson's rule.
The constant $\varphi_{00}$ can be obtained from the Green's function solution of the two-dimensional Poisson equation~\ref{poi2d12} as:
\begin{equation}
\varphi_{00} = \frac{-1}{4\pi\epsilon_0}\int\int log((x^2+y^2)/r_0^2) n(x,y)dxdy
\end{equation}
where $r_0$ is the reference length scale in the open space.
%For a finite real boundary, $r_0$ denotes the distance
%where the potential vanishes.
For a bi-Gaussian distribution, it becomes
\begin{equation}
\varphi_{00} = \frac{-1}{4\pi\epsilon_0} (\frac{1}{2\pi \sigma_x \sigma_y}\int\int \log(x^2+y^2) \exp[-\frac{x^2}{2\sigma_x^2}-\frac{y^2}{2\sigma_y^2}]dxdy - 2\log(r_0))
\end{equation}
The above integral can be rewritten in the polar coordinate system as:
\begin{equation}
\varphi_{00} = \frac{-1}{4\pi\epsilon_0} (\frac{1}{2\pi}\int_0^{\infty}\int_0^{2\pi} \log(r^2(\sigma_x^2\cos^2(\theta)+ \sigma_y^2\sin^2(\theta)))\exp(-\frac{1}{2}r^2)rdrd\theta - 2\log(r_0))
\end{equation}
This integral admits a closed-form solution:
\begin{equation}
\varphi_{00} = \frac{-1}{4\pi\epsilon_0} (\log2-\gamma_{em}+2\log(\frac{\sigma_x+\sigma_y}{2r_0}) )
\end{equation}
where $\gamma_{em}=-\int_0^{\infty}e^{-x} log(x) dx\simeq0.577216$~\cite{math} is the Euler-Mascheroni constant.
\begin{figure}[!htb]
%\begin{figure}[htb]
%   \vspace*{-.5\baselineskip}
   \centering
   \includegraphics*[angle=0,width=230pt]{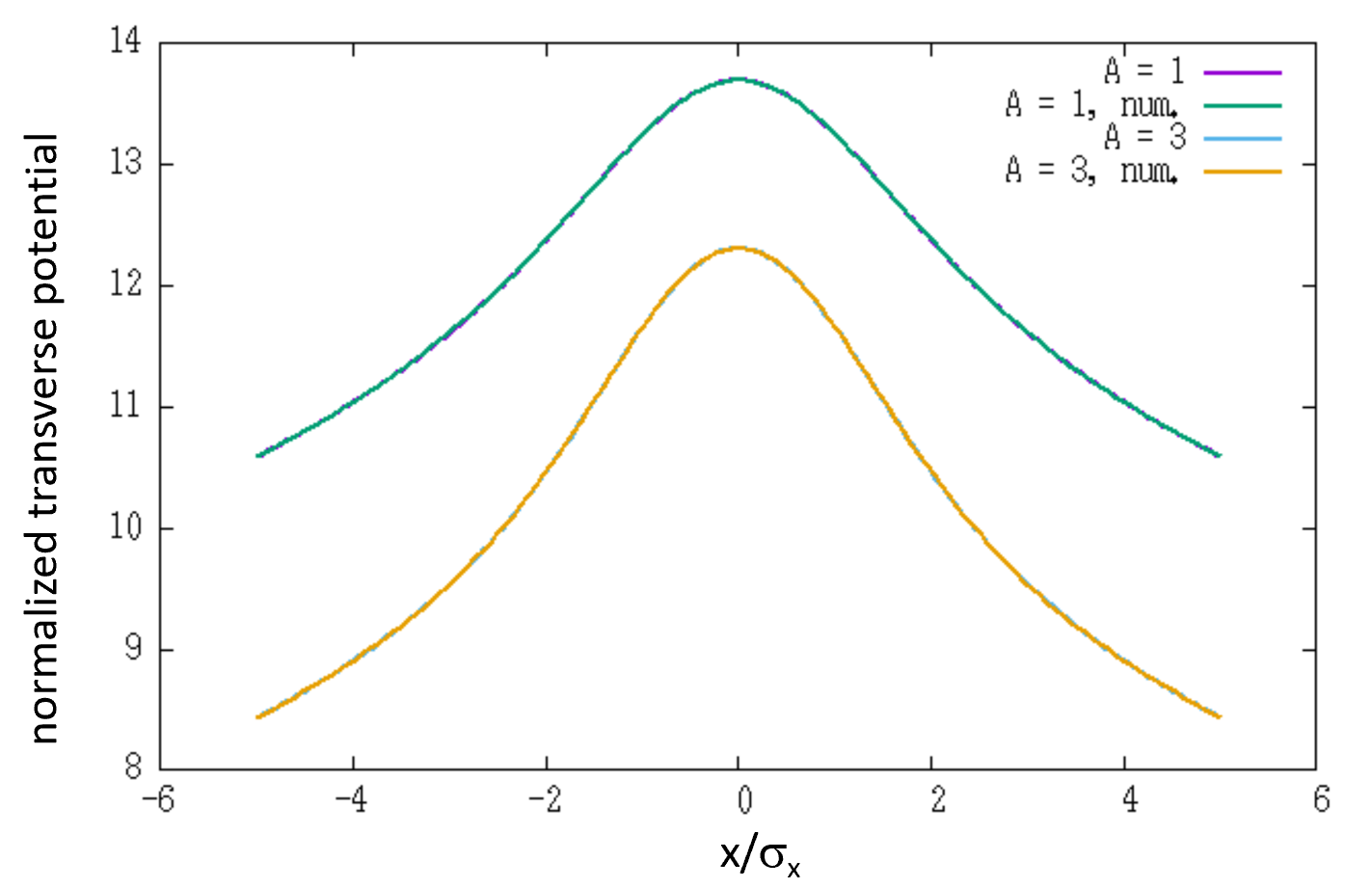}
   \caption{The normalized transverse potential as a function of normalized horizontal distance with
   aspect ratios $1$ (magenta) and $3$ (blue) together with the numerical solutions (green and orange).}
   \label{pot2}
%   \vspace*{-\baselineskip}
\end{figure}
Figure~\ref{pot2} shows the electric potential, $(\varphi(x,0) + \varphi_{00})$, normalized by $1/(4\pi\epsilon_0)$, as a function of the normalized horizontal position for aspect ratios of 1 and 3. Results are presented both from the above integral and from the numerical convolution using the integrated Green's function method described in the next subsection. As shown in the figure, the semi-analytical and numerical solutions overlap, demonstrating excellent agreement between the two methods.

From the above electric potential, we can compute the symplectic
kick in each momentum direction using $H_2$.
The three momentum components are updated after one step $\tau$ as:
\begin{eqnarray}
    p_x(\tau) & = & p_x(0) + \tau K \frac{\lambda(z)}{4\pi\epsilon_0}\int_0^{\infty} dt
    \frac{x\exp[-\frac{x^2}{2(\sigma_x^2+t)}-\frac{y^2}{2(\sigma_y^2+t)}]}{(\sigma_x^2+t)\sqrt{(\sigma_x^2+t)(\sigma_y^2+t) }    }  \\
     p_y(\tau) & = & p_y(0) + \tau K \frac{\lambda(z)}{4\pi\epsilon_0}\int_0^{\infty} dt
    \frac{y\exp[-\frac{x^2}{2(\sigma_x^2+t)}-\frac{y^2}{2(\sigma_y^2+t)}]}{(\sigma_y^2+t)\sqrt{(\sigma_x^2+t)(\sigma_y^2+t) }    }  \\
    p_z(\tau) & = & p_z(0) -\tau K(\varphi(x,y)+\varphi_{00})\frac{\partial \lambda(z)}{\partial z}
\end{eqnarray}
The transverse momenta are typically computed using the Bassetti-Erskine complex error function~\cite{bassetti}, while the update of the longitudinal momentum deviation can be obtained from the integral~\ref{phiapp} and $\varphi_{00}$. Calculating these two integrals for the transverse momenta may be faster than using the complex error function by employing the above $w$ variable and taking advantage of the similarity among the integrands in all three integrals. A Fortran subroutine and a C++ function for computing the integrals in Eqs. 25, 31, and 32 are provided in Appendices A and B.

For a longitudinal Gaussian density distribution $\lambda(z) = \frac{Q}{\sqrt{2\pi}\sigma_z}\exp{(-\frac{z^2}{2\sigma_z^2})}$, the momentum
deviation update will be
\begin{eqnarray}
   p_z(\tau) & = & p_z(0) +\tau K(\varphi(x,y)+\varphi_{00})\frac{Qz}{\sqrt{2\pi}\sigma_z^3}\exp{(-\frac{z^2}{2\sigma_z^2})}
\end{eqnarray}
where $Q$ is the total charge of the beam bunch.

The two-dimensional Poisson equation solution discussed above assumes that the longitudinal variation within a long beam bunch is much smaller than the transverse variation. In the following, we assess the accuracy of this approximation by considering a three-dimensional Gaussian density distribution, for which a semi-analytical solution can be obtained.
For a three-dimensional Gaussian density distribution,
\[
\rho(x, y, z) = \frac{Q}{(2\pi)^{3/2} \sigma_x \sigma_y \sigma_z} \exp\left(-\left(\frac{x^2}{2\sigma_x^2} + \frac{y^2}{2\sigma_y^2} + \frac{z^2}{2\sigma_z^2}\right)\right),
\]
the electric potential in the Poisson equation~\ref{poi3d} can be determined using the same Fourier transform method as in the two-dimensional case.
\begin{equation}
    \phi_{3d}(x,y,z) = \frac{Q}{4\pi\epsilon_0 \sqrt{2\pi}}\int_0^{\infty} dt
    \frac{\exp[-\frac{x^2}{2(\sigma_x^2+t)}-\frac{y^2}{2(\sigma_y^2+t)}-\frac{z^2}{2(\sigma_z^2+t/\gamma^2)}]}{\sqrt{(\sigma_x^2+t)(\sigma_y^2+t)(\sigma_z^2+t/\gamma^2) }    }
    \label{p3dpot}
\end{equation}
%The resulting electric fields ($\bf{E}=-\nabla\phi_{3d}$) in three directions are:
The resulting electric fields, given by $\mathbf{E} = -\nabla\phi_{3d}$, in the three spatial directions are:
\begin{eqnarray}
    E_{x} & = &  \frac{Qx}{4\pi\epsilon_0 \sqrt{2\pi}}\int_0^{\infty} dt
    \frac{\exp[-\frac{x^2}{2(\sigma_x^2+t)}-\frac{y^2}{2(\sigma_y^2+t)}-\frac{z^2}{2(\sigma_z^2+t/\gamma^2)}]}{(\sigma_x^2+t)\sqrt{(\sigma_x^2+t)(\sigma_y^2+t)(\sigma_z^2+t/\gamma^2) }    }  \\
       E_{y} & = &  \frac{Qy}{4\pi\epsilon_0 \sqrt{2\pi}}\int_0^{\infty} dt
    \frac{\exp[-\frac{x^2}{2(\sigma_y^2+t)}-\frac{y^2}{2(\sigma_y^2+t)}-\frac{z^2}{2(\sigma_z^2+t/\gamma^2)}]}{(\sigma_y^2+t)\sqrt{(\sigma_x^2+t)(\sigma_y^2+t)(\sigma_z^2+t/\gamma^2) }    }  \\
       E_{z} & = &  \frac{Qz}{4\pi\epsilon_0 \sqrt{2\pi}}\int_0^{\infty} dt
    \frac{\exp[-\frac{x^2}{2(\sigma_x^2+t)}-\frac{y^2}{2(\sigma_y^2+t)}-\frac{z^2}{2(\sigma_z^2+t/\gamma^2)}]}{(\sigma_z^2+t/\gamma^2)\sqrt{(\sigma_x^2+t)(\sigma_y^2+t)(\sigma_z^2+t/\gamma^2) }    }
\end{eqnarray}
Equation~\ref{p3dpot} can be rewritten using $w$ as the independent variable in the integrand as follows:
\begin{equation}
    \phi_{3d}(x,y,z) = \frac{Q2A}{4\pi\epsilon_0\sqrt{2\pi}\sigma_z}\int_0^{1} dw
    \frac{\exp[-\frac{1}{2}w^2x'^2-\frac{1}{2}w^2y'^2/(A^2+(1-A^2)w^2)-\frac{1}{2}z'^2/(1+\frac{B^2}{w^2}(1-w^2))]}{w\sqrt{(A^2+(1-A^2)w^2)(1+\frac{B^2}{w^2}(1-w^2)) }    }
\end{equation}
where $z'=z/\sigma_z$ and $B=\sigma_x/(\gamma \sigma_z)$.
If $\frac{B}{w}\ll1$, i.e. $1+\frac{B^2}{w^2}(1-w^2) \approx 1$, the above 
equation is reduced to:
\begin{equation}
    \phi_{3d}(x,y,z) \approx \frac{Q2A \exp(-\frac{1}{2}z'^2)}{4\pi\epsilon_0\sqrt{2\pi}\sigma_z}\int_0^{1} dw
    \frac{\exp[-\frac{1}{2}w^2x'^2-\frac{1}{2}w^2y'^2/(A^2+(1-A^2)w^2)]}{w\sqrt{(A^2+(1-A^2)w^2) }    }
\end{equation}
From the definition of $\lambda(z)$ for a longitudinal Gaussian distribution, it can be seen that the above equation corresponds to the zeroth-order two-dimensional potential solution Eq.~\ref{eq25}, without regularization. The error introduced by this zeroth-order two-dimensional approximation can be estimated by the relative difference between the two integrands of potential integrals, given by:
\begin{eqnarray}
    Err(w,B) & = & |\exp(-\frac{1}{2}\frac{B^2(1-w^2)z'^2}{w^2+B^2(1-w^2)})
    \sqrt{1+\frac{B^2}{w^2}(1-w^2)} -1|
    \label{errwb}
\end{eqnarray}
Here, $w$ is between $0$ and $1$. The above function becomes infinite at
$w=0$ except with $B=0$, which suggests a potential large error in the
two-dimensional approximation.
\begin{figure}[!htb]
%\begin{figure}[htb]
%   \vspace*{-.5\baselineskip}
   \centering
   \includegraphics*[angle=0,width=230pt]{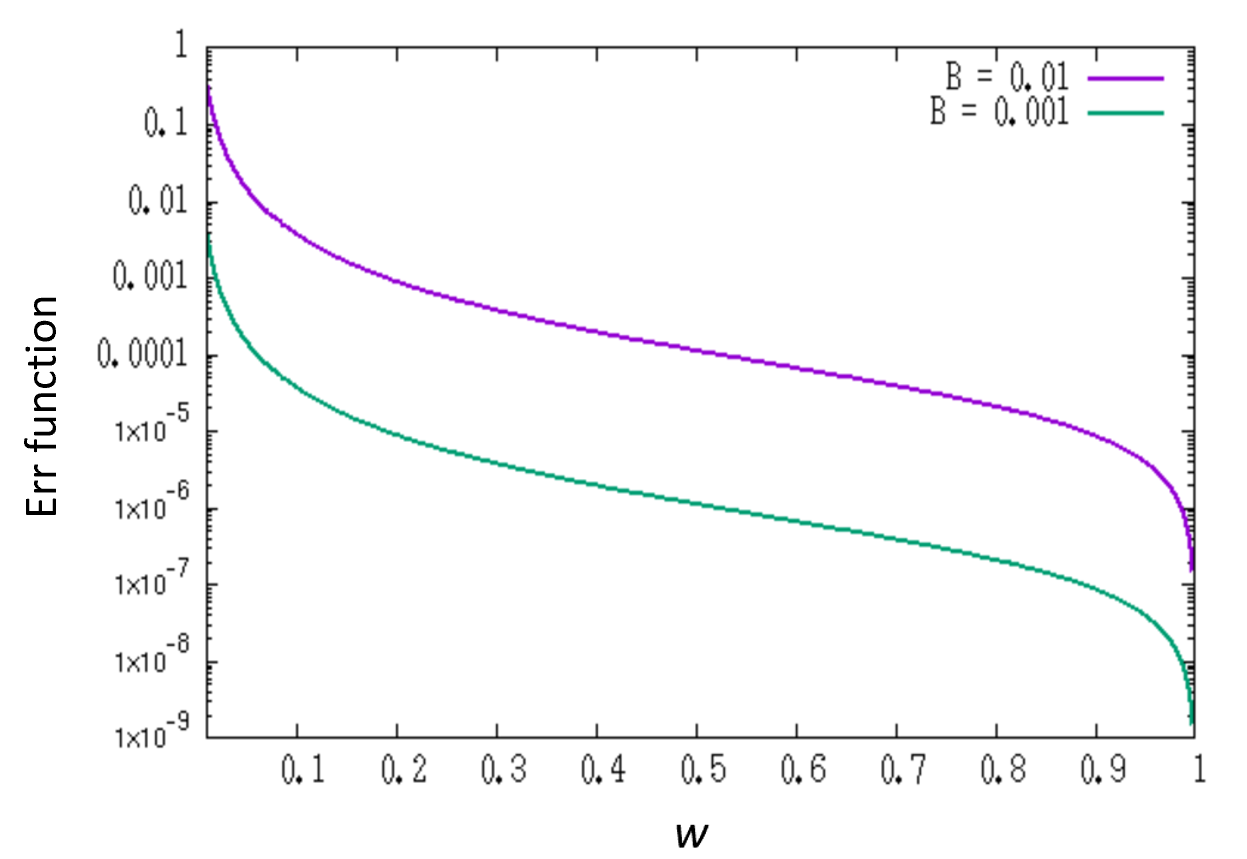}
   \caption{The error defined in Eq.~\ref{errwb} as a function of $w$ with aspect ratios $0.01$ (magenta) and $0.001$ (green) at $(z'=0.5)$.}
   \label{integ2}
%   \vspace*{-\baselineskip}
\end{figure}
Figure~\ref{integ2} shows the above error function as a function of $w$ with two
$B$ values.
It can be seen that the maximum error occurs at the minimum $w$ values and
can be small as long as $B/w$ is small.
For $B/w\ll1$, the above equation can be rewritten as:
\begin{eqnarray}
    Err(w,B) & \approx & |\frac{1}{2}(1-w^2)(1-z'^2)\frac{B^2}{w^2}|
\end{eqnarray}
The maximum error will be
\begin{eqnarray}
    Err(w,B)_{max} & \approx & |\frac{1}{2}(1-w_{min}^2)(1-z'^2)\frac{B^2}{w_{min}^2}|
\end{eqnarray}
Assuming the maximum longitudinal position $z'_{max}=5$, and small $w$ value
$1-w^2\approx 1$, the maximum error becomes
\begin{eqnarray}
    Err(w,B)_{max} & \approx & 12\frac{B^2}{w_{min}^2}
\end{eqnarray}
The choice of $w_{\text{min}}$ and $B$ affects the accuracy of the zeroth-order two-dimensional approximation. Although the minimum value of $w$ is zero, its contribution to the electric fields, as shown below, is also zero. This suggests that the integration over $w$ can begin from a small finite value rather than from zero.
%has little impact on the 
%field values.

\begin{figure}[!htb]
%\begin{figure}[htb]
%   \vspace*{-.5\baselineskip}
   \centering
   \includegraphics*[angle=0,width=230pt]{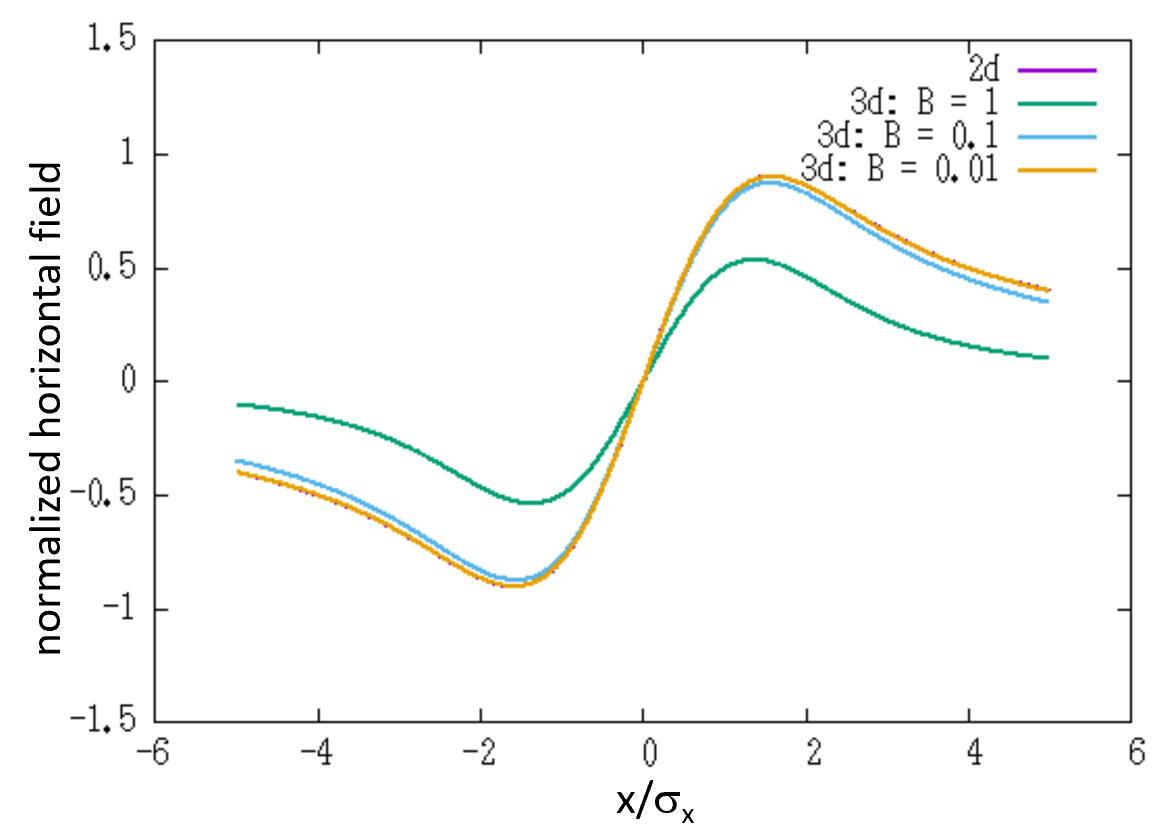}
   \caption{The normalized horizontal electric field as a function of $x/\sigma_x$ from the two dimensional solution with longitudinal aspect ratios $1$ (magenta) and from the three
   dimensional solution with aspect ration $1$ (green), $0.1$ (blue), and $0.01$ (orange) at $(y=z=0)$ and $r_0=1$m.}
   \label{exfld}
%   \vspace*{-\baselineskip}
\end{figure}

The three-dimensional electric fields can be rewritten using the
variable $w$ as:
\begin{eqnarray}
\label{ex}
    E_{x}(x,y,z) & = & \frac{Q2Ax}{4\pi\epsilon_0\sqrt{2\pi}\sigma_z\sigma_x^2}\int_0^{1} dw
    \frac{w^2\exp[-\frac{1}{2}w^2x'^2-\frac{1}{2}w^2y'^2/(A^2+(1-A^2)w^2)-\frac{1}{2}w^2z'^2/(B^2+(1-B^2)w^2)]}{\sqrt{(A^2+(1-A^2)w^2)(B^2+(1-B^2)w^2) }    }  \\
        E_{y}(x,y,z) & = & \frac{Q2Ay}{4\pi\epsilon_0\sqrt{2\pi}\sigma_z\sigma_y^2}\int_0^{1} dw
    \frac{w^2\exp[-\frac{1}{2}w^2x'^2-\frac{1}{2}w^2y'^2/(A^2+(1-A^2)w^2)-\frac{1}{2}w^2z'^2/(B^2+(1-B^2)w^2)]}{(A^2+(1-A^2)w^2)\sqrt{(A^2+(1-A^2)w^2)(B^2+(1-B^2)w^2) }    }  \\
      E_{z}(x,y,z) & = & \frac{Q2Az}{4\pi\epsilon_0\sqrt{2\pi}\sigma_z\sigma_z^2}\int_0^{1} dw
    \frac{w^2\exp[-\frac{1}{2}w^2x'^2-\frac{1}{2}w^2y'^2/(A^2+(1-A^2)w^2)-\frac{1}{2}w^2z'^2/(B^2+(1-B^2)w^2)]}{(B^2+(1-B^2)w^2)\sqrt{(A^2+(1-A^2)w^2)(B^2+(1-B^2)w^2) }    }  
\end{eqnarray}
 \begin{figure}[!htb]
%\begin{figure}[htb]
%   \vspace*{-.5\baselineskip}
   \centering
   \includegraphics*[angle=0,width=230pt]{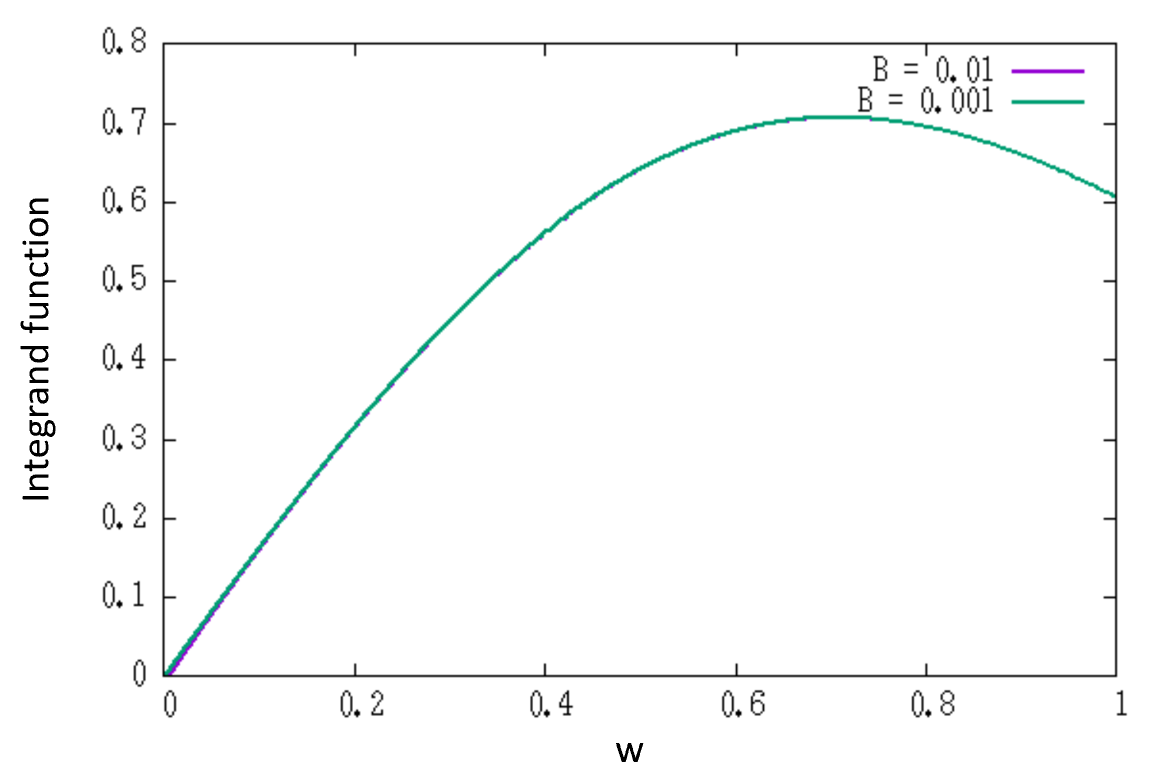}
   \caption{The integrand function in the above integral~\ref{ex} with aspect ratios $0.01$ (magenta) and $0.001$ (green) at $(x'=y'=z'=1)$.}
   \label{integ3}
%   \vspace*{-\baselineskip}
\end{figure}   
Figure~\ref{integ3} shows the integrand of the above integral~\ref{ex} as a function of $w$ for $B=0.01$ and $B=0.001$ (with $x'=y'=z'=1$, $A=1$). It can be seen that the maximum value of the integrand is approximately $0.7$, and the integral itself is on the order of $O(0.1)$. 
For small $w$, the contribution of the integral between $0$ and $w_{\text{min}}$ in the above equations scales as $O(w_{\text{min}}^3)$. For $w_{\text{min}}=0.01$, this results in an integral of order $O(10^{-6})$. Consequently, setting $w_{\text{min}}=0.01$ leads to a loss of accuracy in the field calculation of $O(10^{-5})$.
As an illustration, we consider a three-dimensional Gaussian bunch with $\sigma_x = \sigma_y = \sigma_z = 1$~mm and various values of $\gamma$. Figure~\ref{exfld} shows the horizontal electric field (normalized by $\frac{Q}{4\pi \epsilon_0 \sqrt{2\pi} \sigma_z}$) as a function of $x/\sigma_x$ at $y = z = 0$ for $\gamma = 1$, $\gamma = 10$, and $\gamma = 100$, comparing the results from the above two-dimensional zeroth-order solution and the full three-dimensional solution.
 \begin{figure}[!htb]
%\begin{figure}[htb]
%   \vspace*{-.5\baselineskip}
   \centering
   \includegraphics*[angle=0,width=230pt]{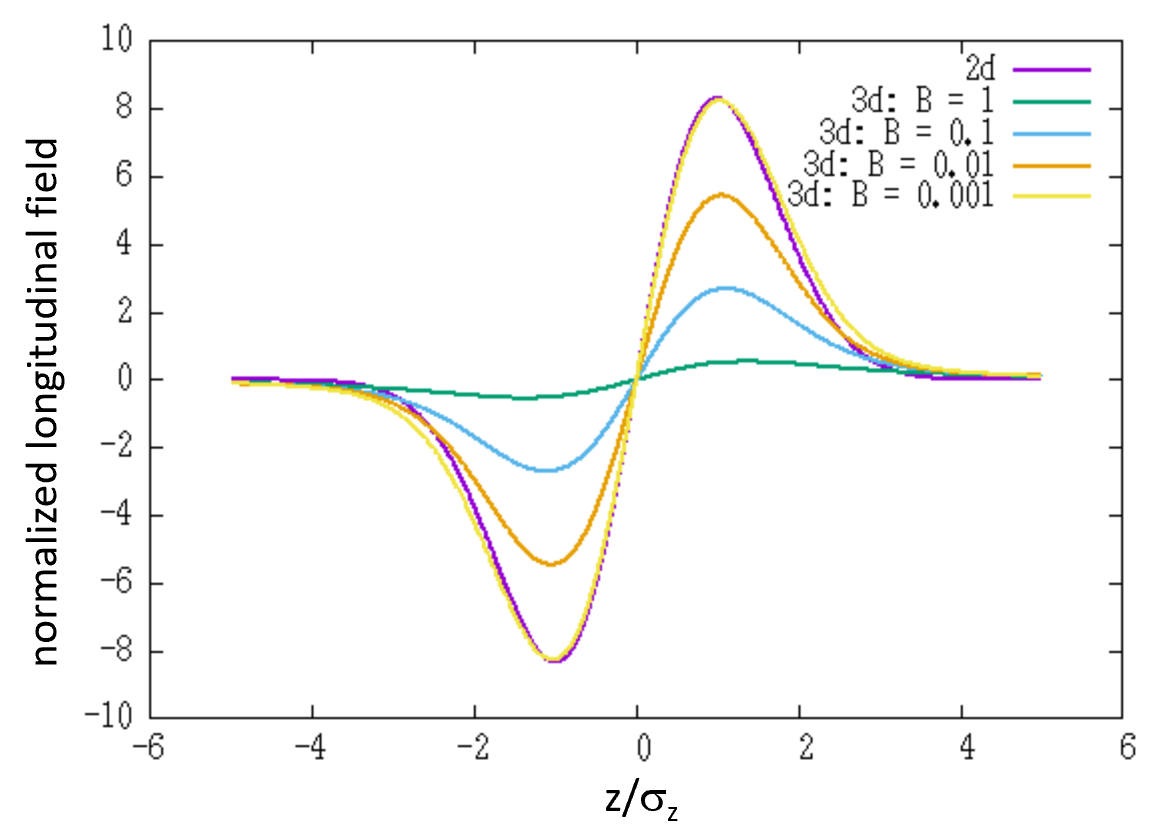}
   \caption{The normalized longitudinal electric field as a function of $z/\sigma_z$ from the two dimensional solution with longitudinal aspect ratios $1$ (magenta) and from the three
   dimensional solution with aspect ration $1$ (green), $0.1$ (blue), $0.01$ (orange), and $0.001$ (yellow) at $(x=y=0)$ and $r_0=1$m.}
   \label{ezfld}
%   \vspace*{-\baselineskip}
\end{figure}   

Figure~\ref{ezfld} shows the normalized longitudinal electric field as a function of $z/\sigma_z$ at $x = y = 0$ for $\gamma = 1$, $10$, $100$, and $1000$, comparing the results from the above two-dimensional zeroth-order solution and the three-dimensional solution. It can be seen that as the longitudinal aspect ratio $B$ decreases, the three-dimensional solution approaches the two-dimensional solution (independent of $\gamma$) for both the horizontal and longitudinal electric fields. This indicates that the two-dimensional solution provides a good approximation to the three-dimensional solution when the transverse-to-longitudinal aspect ratio is small, corresponding to a very long bunch or a high-energy beam with a large $\gamma$ factor.

In the above example, the reference length scale $r_0$ is set to $1~\mathrm{m}$.
For the 3D open-space geometry, $r_0$ is chosen by minimizing the discrepancy
$D(r_0)$ between the on-axis longitudinal field $E_z$ from the 2.5D model and that
from the full 3D model:
\begin{equation}
D(r_0)=\int_{-\infty}^{\infty}
\left(E_z^{2.5D}(0,0,z;r_0)-E_z^{3D}(0,0,z)\right)^2\,dz.
\end{equation}
Here,
\begin{align}
E_z^{2.5D}(0,0,z;r_0)
&=\varphi_{00}(r_0)\frac{Qz}{\sqrt{2\pi}\sigma_z^3}
\exp\!\left(-\frac{z^2}{2\sigma_z^2}\right),\\
E_z^{3D}(0,0,z)
&=\frac{Qz}{4\pi\epsilon_0\sqrt{2\pi}}
\int_0^{\infty} dt\,
\frac{\exp\!\left[-\frac{z^2}{2(\sigma_z^2+t/\gamma^2)}\right]}
{(\sigma_z^2+t/\gamma^2)\sqrt{(\sigma_x^2+t)(\sigma_y^2+t)(\sigma_z^2+t/\gamma^2)}}.
\end{align}

After minimizing $D(r_0)$ with respect to $r_0$, one obtains
\begin{equation}
\log(r_0)=
\log\!\left(\frac{\sqrt{2}(\sigma_x+\sigma_y)}{2}\right)
-\frac{1}{2}\gamma_{em}
+\sqrt{2}\sigma_z^3\gamma^3
\int^{\infty}_0 dt\,
\frac{1}{(2\sigma_z^2\gamma^2+t)^{3/2}\sqrt{(\sigma_x^2+t)(\sigma_y^2+t)}}.
\end{equation}

The integral appearing above is one of the Carlson elliptic integrals,
which can be evaluated numerically~\cite{math}.
For a round beam with $\sigma_x=\sigma_y=\sigma$, an analytical expression can be derived:
\begin{equation}
\log(r_0)=
\log(\sqrt{2}\sigma)
-\frac{1}{2}\gamma_{em}
+\frac{2^{3/2}\sigma_z^3\gamma^3}{(2\sigma_z^2\gamma^2-\sigma^2)^{3/2}}
\Bigg[
-\sqrt{1-\frac{\sigma^2}{2\sigma_z^2\gamma^2}}
-\log\!\left(\frac{\sigma}{\sqrt{2}\sigma_z \gamma}\right)
+\log\!\left(1+\sqrt{1-\frac{\sigma^2}{2\sigma_z^2\gamma^2}}\right)
\Bigg].
\end{equation}

For a long bunch in the beam frame, $\sigma \ll \gamma \sigma_z$,
the above expression reduces to
\begin{equation}
r_0 = 4e^{(-0.5\gamma_{em}-1)}\gamma\sigma_z  \sim 1.1\,\gamma \sigma_z.
\end{equation}

For the round perfectly conducting pipe geometry, the on-axis potential of a round Gaussian beam can be approximated as
\begin{equation}
\varphi_{00}\approx\frac{-1}{4\pi\epsilon_0}\left(\log 2-\gamma_{em}+2\log\!\left(\frac{\sigma}{a}\right)\right),
\end{equation}
where $a$ is the pipe radius.
This suggests that, in this scenario, the reference length scale $r_0$ may be chosen as the pipe radius, i.e., $r_0=a$.

For reference, Fortran and C++ implementations of the above integrals for computing the three-dimensional electric fields from the three-dimensional Gaussian distribution are provided in Appendices~C and~D.

If the longitudinal density distribution $\lambda(z)$ is not a Gaussian function, it can be decomposed as a sum of Gaussian wavelets, i.e.,
\begin{equation}
    \lambda(z) = \frac{Q}{\sqrt{2\pi}} \sum_{i=1}^N \frac{1}{\sigma_{z_i}} \exp\left(-\frac{1}{2} \frac{(z - z_i)^2}{\sigma_{z_i}^2}\right),
\end{equation}
where $z_i$ and $\sigma_{z_i}$ are the centroid and standard deviation of the $i$th wavelet, respectively, and $N$ is the total number of wavelets. 

The electric potential solution to the three-dimensional Poisson equation, for a transverse Gaussian distribution and the longitudinal distribution given above, can be expressed as:
\begin{equation}
    \phi_{3d}(x,y,z) = \frac{Q}{4\pi\epsilon_0 \sqrt{2\pi}}\int_0^{\infty} dt
    \frac{\exp[-\frac{x^2}{2(\sigma_x^2+t)}-\frac{y^2}{2(\sigma_y^2+t)}]}{\sqrt{(\sigma_x^2+t)(\sigma_y^2+t)}}
    (\sum_{i=1}^N\frac{\exp[-\frac{(z-z_i)^2}{2(\sigma_{z_i}^2+t/\gamma^2)}]}{\sqrt{(\sigma_{z_i}^2+t/\gamma^2) }}    )
    \label{p3dpot2}
\end{equation}
The resulting electric fields can be written as:
\begin{eqnarray}
    E_{x}(x,y,z) & = &\frac{Qx}{4\pi\epsilon_0 \sqrt{2\pi}}\int_0^{\infty} dt
    \frac{\exp[-\frac{x^2}{2(\sigma_x^2+t)}-\frac{y^2}{2(\sigma_y^2+t)}]}{(\sigma_x^2+t)\sqrt{(\sigma_x^2+t)(\sigma_y^2+t)}}
    (\sum_{i=1}^N\frac{\exp[-\frac{(z-z_i)^2}{2(\sigma_{z_i}^2+t/\gamma^2)}]}{\sqrt{(\sigma_{z_i}^2+t/\gamma^2) }}    )  \\
    E_{y}(x,y,z) & = &\frac{Qy}{4\pi\epsilon_0 \sqrt{2\pi}}\int_0^{\infty} dt
    \frac{\exp[-\frac{x^2}{2(\sigma_x^2+t)}-\frac{y^2}{2(\sigma_y^2+t)}]}{(\sigma_y^2+t)\sqrt{(\sigma_x^2+t)(\sigma_y^2+t)}}
    (\sum_{i=1}^N\frac{\exp[-\frac{(z-z_i)^2}{2(\sigma_{z_i}^2+t/\gamma^2)}]}{\sqrt{(\sigma_{z_i}^2+t/\gamma^2) }}    )  \\
     E_{z}(x,y,z) & = &\frac{Q}{4\pi\epsilon_0 \sqrt{2\pi}}\int_0^{\infty} dt
    \frac{\exp[-\frac{x^2}{2(\sigma_x^2+t)}-\frac{y^2}{2(\sigma_y^2+t)}]}{\sqrt{(\sigma_x^2+t)(\sigma_y^2+t)}}
    (\sum_{i=1}^N\frac{(z-z_i)\exp[-\frac{(z-z_i)^2}{2(\sigma_{z_i}^2+t/\gamma^2)}]}{(\sigma_{z_i}^2+t/\gamma^2)\sqrt{(\sigma_{z_i}^2+t/\gamma^2) }}    )  
    \label{p3fld3}
\end{eqnarray}

\subsection{Space-charge solver with an arbitrary density distribution}

 The semi-analytical expression for space-charge fields derived from a Gaussian distribution can facilitate rapid tracking and parameter space exploration~\cite{ingo21,adrian22}. However, for accurate predictions of emittance growth or beam loss, particularly in regions near resonances or in the presence of machine errors, the particle density distribution may deviate significantly from a Gaussian function, necessitating the use of a direct numerical space-charge solver.

For an arbitrary transverse charge density distribution, numerical methods must be employed. Various numerical techniques can effectively solve the two-dimensional Poisson equation under different boundary conditions. For transverse open boundary conditions, an efficient approach is to use the Green's function method combined with the Fast Fourier Transform (FFT) to perform cyclic summation on a two-dimensional grid. The Green's function solution to two-dimensional Poisson's equation~\ref{poi2d12} can be expressed as:
\begin{equation}
\varphi(x,y) = \frac{1}{2\pi\epsilon_0}\int\int G(x-x',y-y') n(x',y')dx'dy'
\end{equation}
where
\begin{equation}
G(x,y) = -\frac{1}{2} log(x^2+y^2)
\end{equation}
The above convolution can be computed
numerically on a grid using an integrated Green's function
method~\cite{bb3d}.
%if the  spatial variation of the density
%is smooth.
%\begin{equation}
%\phi(x,y) = \sum\frac{-1}{4\pi\epsilon_0}\int\int log((x-x')^2+(y-y')^2) n(x',y')dx'dy'
%\end{equation}
\begin{eqnarray}
{\varphi}(x_i,y_j) & = & \frac{1}{2 \pi \epsilon_0} 
\sum_{i'=1}^{N_x} \sum_{j'=1}^{N_y} {\bar G}(x_i-x_{i'},y_j-y_{j'}) 
n(x_{i'},y_{j'})
\end{eqnarray}
where
\begin{eqnarray}
{\bar G}(x_i-x_{i'},y_j-y_{j'}) & = & \int_{x_{i'}-h_x/2}^{x_{i'}+h_x/2} dx'
\int_{y_{j'}-h_y/2}^{y_{j'}+h_y/2} dy' G(x_i-x',y_j-y')
\end{eqnarray}
This integration can be done analytically using the indefinite integral:
\begin{equation}
\int \int \ln (x^2+y^2)dxdy = - 3xy + x^2 \arctan(y/x) + y^2 \arctan(x/y)
+ xy \ln (x^2 + y^2)
\end{equation}
The cyclic summation for the electric potential described above can be computed efficiently using an FFT~\cite{hockney}.
%From the two-dimensional potential computed on the grid, a spline shape function $S$ can be used to obtain the space-charge potential and its derivatives with respect to $x$ and $y$. The one-step momentum update from the space-charge map can then be expressed more concisely as:
From the two-dimensional potential computed on the grid, a spline shape function $S$ can be used to obtain the space-charge potential and its derivatives with respect to $x$ and $y$. The one-step momentum update from the space-charge map can then be expressed concisely as:
\begin{eqnarray}
		p_{x}(\tau) & = & p_{x}(0) -
	\tau {K} \lambda(z)\sum_i \sum_j \frac{\partial S(x_i-x)}{\partial x} S(y_j-y) 
\varphi(x_i,y_j)	\nonumber \\
	p_{y}(\tau) & = & p_{y}(0) -
	\tau {K} \lambda(z) \sum_i \sum_j S(x_i-x)\frac{\partial S(y_j-y)}{\partial y} 
	\varphi(x_i,y_j) \\
    p_{z}(\tau) & = & p_{z}(0) -
	\tau {K} \frac{\partial \lambda(z)}{\partial z} \sum_i \sum_j S(x_i-x) S(y_j-y) 
	\varphi(x_i,y_j)
\end{eqnarray}

Next, we consider the case where the electric potential is confined within a rectangular, perfectly conducting pipe. The charge density distribution, $\rho(x, y, z)$, is still assumed to be separable into a product of a transverse distribution, $n(x, y)$, and a longitudinal distribution, $\lambda(z)$.
The boundary conditions for the electric potential $\varphi(x,y)$ inside the rectangular 
perfectly conducting pipe are:
\begin{eqnarray}
	\label{bc1}
\varphi(x=0,y) & = & 0  \\
\varphi(x=a,y) & = & 0  \\
\varphi(x,y=0) & = & 0  \\
\varphi(x,y=b) & = & 0  
	\label{boundary}
\end{eqnarray}
where $a$ is the horizontal width of the pipe and $b$ is the vertical width
of the pipe. 

Given the boundary conditions in Eqs.~\ref{bc1}-\ref{boundary}, the electric potential $\varphi$ and the
source term $n(x,y)$ can be approximated using two sine functions as:%~\cite{gottlieb,fornberg,boyd,qiang1,qiang2}:
\begin{eqnarray}
	n(x,y)  = \sum_{l=1}^{N_l}\sum_{m=1}^{N_m} n^{lm} \sin(\alpha_l x) \sin(\beta_m y) \\
	\varphi(x,y)  =  \sum_{l=1}^{N_l}\sum_{m=1}^{N_m} \varphi^{lm} \sin(\alpha_l x) \sin(\beta_m y) 
\end{eqnarray}
where
%\begin{small}
\begin{eqnarray}
\label{rholm}
n^{lm}  = \frac{4}{ab}\int_0^a\int_0^b n(x,y) \sin(\alpha_l x) \sin(\beta_m y) \ dx dy \\
\varphi^{lm}  = \frac{4}{ab}\int_0^a\int_0^b \varphi(x,y) \sin(\alpha_l x) \sin(\beta_m y) \ dx dy
\end{eqnarray}
%\end{small}
where $\alpha_l=l\pi/a$ and $\beta_m = m \pi/b$.
The above approximation
follows the numerical spectral Galerkin method since each basis function
satisfies the boundary conditions on the wall.
%~\cite{gottlieb,boyd,forn}. 
For a smooth function,
this spectral approximation has an accuracy whose numerical error
scales as $O(\exp(-cN))$ with 
$c>0$, where $N$ is the number of the basis function (i.e. mode number in each
dimension) used in the approximation~\cite{gottlieb,fornberg}.
By substituting above expansions into the
Poisson Eq.~\ref{poi2d12} and making use of the orthonormal condition of the sine functions,
we obtain
\begin{eqnarray}
	\varphi^{lm} & = & \frac{n^{lm}}{\epsilon_0\gamma_{lm}^2}
	\label{odelm}
\end{eqnarray}
where $\gamma_{lm}^2 = \alpha_l^2 + \beta_m^2$. 
The zeroth-order approximate solution of potential $\phi_0$ can be written as:
\begin{equation}
\phi_0(x,y,z) = \varphi(x,y)\lambda(z)
\end{equation}
From the above potential, the symplectic momentum kicks after each step
due to the space-charge effects are:
\begin{eqnarray}
    p_x(\tau) & = & p_x(0) - \tau K \lambda(z) \sum_{l=1}^{N_l}\sum_{m=1}^{N_m} \varphi^{lm} \alpha_l\cos(\alpha_l x) \sin(\beta_m y)    \\
     p_y(\tau) & = & p_y(0) - \tau K \lambda(z)
      \sum_{l=1}^{N_l}\sum_{m=1}^{N_m} \varphi^{lm} \beta_m \sin(\alpha_l x) \cos(\beta_m y)  \\
    p_z(\tau) & = & p_z(0) -\tau K    \frac{\partial \lambda(z)}{\partial z} \sum_{l=1}^{N_l}\sum_{m=1}^{N_m} \varphi^{lm} \sin(\alpha_l x) \sin(\beta_m y) 
\end{eqnarray}

\begin{figure}[!htb]
%\begin{figure}[htb]
%   \vspace*{-.5\baselineskip}
   \centering
   \includegraphics*[angle=0,width=230pt]{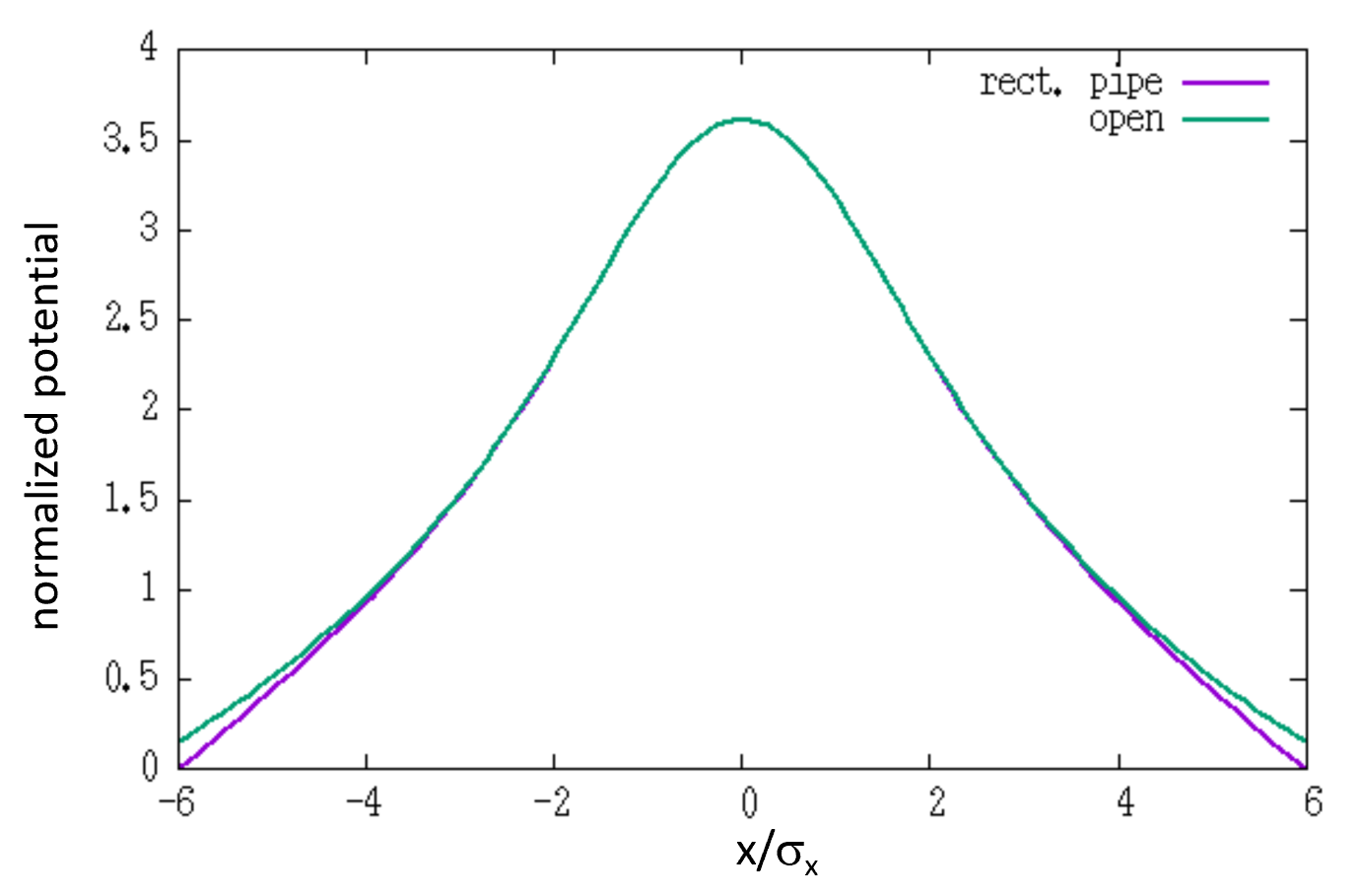}
   \caption{The normalized electric potential as a function horizontal location from
   the spectral solver inside a rectangular pipe and from the Green's function solver with open
   boundary conditions.}
   \label{potnum}
%   \vspace*{-\baselineskip}
\end{figure}

As a test of the above space-charge solver, we assumed a two-dimensional Gaussian density 
distribution with $\sigma_x = \sigma_y = 1$~mm and computed the electric potential using both the
Green's function method with open boundary conditions and the spectral method with a 
perfectly conducting rectangular pipe of $6\times6$~mm aperture size.
Figure~\ref{potnum} shows the normalized electric potential (normalized by $\frac{\lambda(z)}{4\pi \epsilon_0}$)
as a function of horizontal position, calculated using both the Green's function method and the spectral method.
Here, the electric potential from the Green's function method is shifted to match the maximum of
the spectral solution inside the pipe. 
A Gaussian distribution extends to infinity,
the spectral approach necessarily truncates the distribution at its boundary. The potential will be shielded from the hypothetical
charges beyond the boundary resulting in different maximum potential
values from the open boundary solution. 
%This explains why the maximum potential
%values must differ.
After shifting, the two solutions agree well within
the core of the distribution. The slight difference near the edge of the computational domain is
noticeable and is due to the different boundary conditions used. For the spectral solution, the
electric potential must be zero on the boundary as required by the boundary condition.
For the Green's function solution, there is no such constraint, so the electric potential at the
edge of the computational domain is not necessarily zero.

For a pipe with a circular cross section, the two-dimensional Poisson equation~\ref{poi2d12} can be written in cylindrical coordinates as:
\begin{equation}
\frac{\partial^2 \varphi}{\partial r^2} +
\frac{1}{r}\frac{\partial \varphi}{\partial r} + \frac{1}{r^2}
\frac{\partial^2 \varphi}{\partial \theta^2}  = - \frac{n(r,\theta)}{\epsilon_0}
\end{equation}
%Here, $\varphi$ denotes the dimensionless electrostatic potential, 
%$\rho$ the dimensionless charge
%density function, $r$ and $z$ the dimensionless radial 
%and longitudinal distance. 
The boundary conditions for the potential are:
\begin{eqnarray}
\varphi(r=a,\theta,z) & = & 0  \\
\varphi(r,\theta + 2\pi,z) & = & \varphi(r,\theta,z)  
\label{poi2d3}
\end{eqnarray}
where $a$ is the pipe radius.

The periodic boundary condition for the potential along the $\theta$ direction suggests the use of complex exponential eigenfunctions in that direction. In the radial direction, Bessel functions serve as appropriate eigenfunctions for a round conducting pipe. Hence, we can approximate the potential $\varphi$ and the source term $n(r, \theta)$ as follows~\cite{qiang2001}:
\begin{eqnarray}
n(r,\theta) &=& \sum^{N_m/2-1}_{m=-N_m/2} \sum^{N_l}_{l=1} n^{lm}J_m(\gamma_{lm}r)\exp(-i m \theta) \\
\label{phibess}
\varphi(r,\theta) &=& \sum^{N_m/2-1}_{m=-N_m/2} \sum^{N_l}_{l=1} \varphi^{lm}J_m(\gamma_{lm}r)\exp(-i m \theta),
\end{eqnarray}
where $\gamma_{lm}$ is a solution of 
\begin{equation}
J_m(\gamma_{lm}a ) = 0.
\end{equation}
The $n^{lm}$ and $\varphi^{lm}$ are determined from
\begin{eqnarray}
n^{lm} & = & \frac{1}{\pi J'^2_m(\gamma_{lm})} \int^{2 \pi}_0
                   \int^a_0 n(r,\theta) \exp(im \theta)
                    r J_m(\gamma_{lm}r) dr d\theta \\
\varphi^{lm} & = & \frac{1}{\pi J'^2_m(\gamma_{lm})} \int^{2 \pi}_0
                   \int^a_0 \varphi(r,\theta) \exp(im \theta) 
                    r J_m(\gamma_{lm}r) dr d\theta.
\end{eqnarray}
%For the first iteration, 
Substituting the above expansion into the Poisson equation and multiplying by $\exp(im \theta) r J_m(\gamma_{lm}r)$ and integrating
from $0$ to $2 \pi$ and $0$ to $a$, we obtain
\begin{equation}
\varphi^{lm}
= \frac{n^{lm}}{\epsilon_0\gamma^{2}_{lm}}
\end{equation}

From the $\varphi^{lm}$, the zeroth-order approximate of potential $\phi_0$ can be obtained using Eq.~\ref{phibess}.
Given the electric potential
$\phi_0(r,\theta,z) =\varphi(r,\theta)\lambda(z)$,
the symplectic momentum kicks after one step $\tau$
due to the space-charge effects are:
\begin{eqnarray}
    p_x(\tau) & = & p_x(0) - \tau K \lambda(z) \sum^{N_m/2-1}_{m=-N_m/2} \sum^{N_l}_{l=1} \varphi^{lm}(\gamma_{lm}J'_m(\gamma_{lm}r) \frac{x}{r}+i\ m\frac{y}{r^2}J_m(\gamma_{lm}r))\exp(-i m \theta),   \\
    p_y(\tau) & = & p_y(0) - \tau K \lambda(z) \sum^{N_m/2-1}_{m=-N_m/2} \sum^{N_l}_{l=1} \varphi^{lm}(\gamma_{lm}J'_m(\gamma_{lm}r) \frac{y}{r}-i\ m\frac{x}{r^2}J_m(\gamma_{lm}r))\exp(-i m \theta),   \\
    p_z(\tau) & = & p_z(0) -\tau K    \frac{\partial \lambda(z)}{\partial z} \sum^{N_m/2-1}_{m=-N_m/2} \sum^{N_l}_{l=1} \varphi^{lm}J_m(\gamma_{lm}r)\exp(-i m \theta) 
\end{eqnarray}
%$\mathcal{Re}$

\section{Two-and-a-half dimensional space-charge solver in a circular accelerator}
In this section, we consider the space-charge solver for a long bunch in a circular accelerator. In this case, we employ the Frenet-Serret coordinate system to formulate Poisson's equation, assuming a curvature $h$ in the horizontal plane.
\begin{equation}
    \frac{1}{1+hx}\frac{\partial}{\partial x}((1+hx)\frac{\partial \phi}{\partial x})+\frac{\partial^2 \phi}{\partial y^2} +\frac{1}{\gamma^2}\frac{1}{1+hx}\frac{\partial}{\partial z}(\frac{1}{1+hx}\frac{\partial \phi}{\partial z}) = -\frac{\rho}{\epsilon_0}
\end{equation}
For a constant curvature, the above equation can be rewritten as:
\begin{equation}
   \frac{\partial^2 \phi}{\partial x^2} + \frac{h}{1+hx}\frac{\partial \phi}{\partial x}+ \frac{\partial^2 \phi}{\partial y^2} + \frac{1}{\gamma^2 (1+hx)^2}\frac{\partial^2 \phi}{\partial z^2} = -\frac{\rho}{\epsilon_0}
\end{equation}
%In most circular accelerators, the horizontal beam size
%is much smaller than the bending radius, i.e. $hx \ll 1$.
In most circular accelerators, the horizontal beam size is much smaller than the bending radius, i.e., $h x \ll 1$. By noting that 
$\frac{h}{1+hx}\frac{\partial \phi}{\partial x} \ll \frac{\partial^2 \phi}{\partial x^2}+\frac{\partial^2 \phi}{\partial y^2} $,
and $\frac{1}{\gamma^2 (1+hx)^2}\frac{\partial^2 \phi}{\partial z^2} \ll \frac{\partial^2 \phi}{\partial x^2}+\frac{\partial^2 \phi}{\partial y^2}$, the equation above can be solved using an iterative procedure,
which is equivalent to a perturbative approach.
%The solution of the above Poisson equation can be written
%as
%\begin{equation}
%    \phi(x,y,z) = \phi_0(x,y,z) + h \phi_1(x,y,z) + h^2 \phi_2(x,y,z)+\cdots
%\end{equation}
%where 
\begin{eqnarray}
\frac{\partial^2 \phi_0}{\partial x^2} +
\frac{\partial^2 \phi_0}{\partial y^2} &  =  & - \frac{\rho(x,y,z)}{\epsilon_0}  \\
\label{poi2dh}
\frac{\partial^2 \phi_1}{\partial x^2} +
\frac{\partial^2 \phi_1}{\partial y^2} &  =  & - \frac{\rho(x,y,z)}{\epsilon_0}  
- \frac{h}{1+hx}\frac{\partial \phi_0}{\partial x} - \frac{1}{\gamma^2 (1+hx)^2}\frac{\partial^2 \phi_0}{\partial z^2}\\
\frac{\partial^2 \phi_2}{\partial x^2} +
\frac{\partial^2 \phi_2}{\partial y^2} &  =  & - \frac{\rho(x,y,z)}{\epsilon_0}  
- \frac{h}{1+hx}\frac{\partial \phi_1}{\partial x} - \frac{1}{\gamma^2 (1+hx)^2}\frac{\partial^2 \phi_1}{\partial z^2}\\
\vdots & = & \vdots  \nonumber
\label{poi3d1}
\end{eqnarray}
The above equations provide the correction to the space-charge potential due to the finite curvature in a circular accelerator. For the first-order correction, Eq.~\ref{poi2dh} can be written as:
\begin{eqnarray}
\frac{\partial^2 \phi_1}{\partial x^2} +
\frac{\partial^2 \phi_1}{\partial y^2} &  =  & - \frac{\rho(x,y,z)}{\epsilon_0}  
- h\frac{\partial \phi_0}{\partial x} 
\end{eqnarray}
For a transverse Gaussian density distribution, and using the zeroth-order solution $\phi_0$ of Eq.~\ref{phi0} from Section~III.A, the solution to the first-order equation above can be obtained as:
\begin{eqnarray}
    \phi_1(x,y,z) & = & \phi_0(x,y,z) - h\frac{\lambda(z) x}{8 \pi \epsilon_0}\int_0^{\infty} t dt
    \frac{\exp[-\frac{x^2}{2(\sigma_x^2+t)}-\frac{y^2}{2(\sigma_y^2+t)}]-1}{(\sigma_x^2+t)\sqrt{(\sigma_x^2+t)(\sigma_y^2+t) }    }
    \label{sol2}
\end{eqnarray}

Here, we have included a term in the potential that depends linearly on $x$ to eliminate the divergence present in the original integral. A derivation of the solution above is provided in Appendix~E.

\begin{figure}[!htb]
%\begin{figure}[htb]
%   \vspace*{-.5\baselineskip}
   \centering
   \includegraphics*[angle=0,width=230pt]{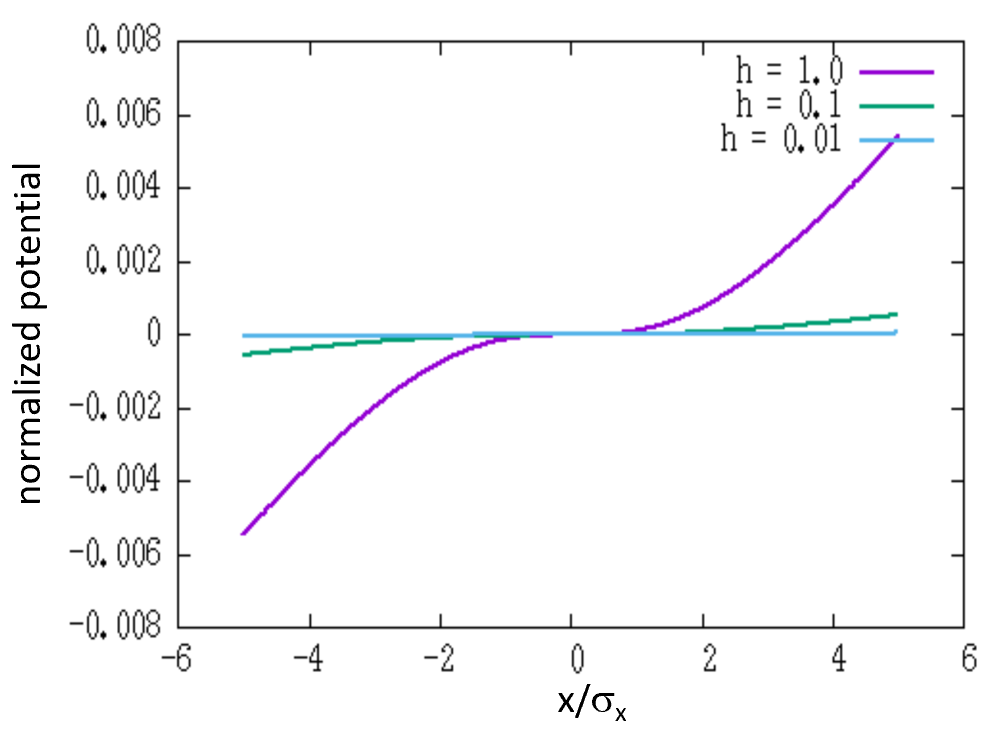}
   \caption{The normalized potential as a function of normalized horizontal distance with
   curvature $1$/m (magenta), $0.1$/m (green), and $0.01$/m (blue).}
   \label{pot2cor}
%   \vspace*{-\baselineskip}
\end{figure}

As an illustration of the effect of curvature correction, we used a Gaussian density distribution with $\sigma_x = \sigma_y = 1$~mm, and considered several curvature values: $1.0$, $0.1$, and $0.01$/m. Figure~\ref{pot2cor} shows the normalized horizontal electric potential correction from Eq.~\ref{sol2} as a function of the horizontal position at $y = z = 0$. For comparison, the electric potential in the straight system, $\phi_0(x, 0, 0)$, is shown in Fig.~\ref{pot2} with a maximum value of about $13.7$. 
Compared with the solution for the straight system, this correction is several orders of magnitude smaller and is proportional to $h\sigma_x$.
%where $R$ is the radius of the circular accelerator. 
For a typical circular accelerator with a radius greater than $100$ meters and a beam size on the order of millimeters, this correction will be $O(10^{-5})$. Therefore, the effect of curvature is likely negligible.

Next, we consider the case of a pipe with a circular cross section. The three-dimensional Poisson equation can then be rewritten in transverse polar coordinates as:
\begin{eqnarray}
 \nabla_{\perp}^2 \phi + h
\frac{1}{1+h r \cos(\theta)} (
\cos(\theta)\frac{\partial \phi}{\partial r} -
\frac{\sin(\theta)}{r} \frac{\partial \phi}{\partial \theta} ) +
 \frac{1}{\gamma^2 (1+h r \cos(\theta))^2}\frac{\partial^2 \phi}{\partial z^2} & = & -\frac{\rho}{\epsilon_0} 
\end{eqnarray}
This equation can be solved using the same iterative procedure:
%that is equivalent to perturbative solution:
\begin{eqnarray}
\nabla_{\perp}^2 \phi_0 & = &  -\frac{\rho}{\epsilon_0} \\
\nabla_{\perp}^2 \phi_1 & = & -\frac{\rho}{\epsilon_0} -h
\frac{1}{1+hr \cos(\theta)} (
\cos(\theta)\frac{\partial \phi_0}{\partial r} -
\frac{\sin(\theta)}{r} \frac{\partial \phi_0}{\partial \theta} ) +   \frac{1}{\gamma^2 (1+h r \cos(\theta))^2}\frac{\partial^2 \phi_0}{\partial z^2}\\ 
\nabla_{\perp}^2 \phi_2 & = & -\frac{\rho}{\epsilon_0} -h
\frac{1}{1+hr \cos(\theta)} (
\cos(\theta)\frac{\partial \phi_1}{\partial r} -
\frac{\sin(\theta)}{r} \frac{\partial \phi_1}{\partial \theta} ) +   \frac{1}{\gamma^2 (1+h r \cos(\theta))^2}\frac{\partial^2 \phi_1}{\partial z^2}\\ 
\vdots & = & \vdots  \nonumber
\end{eqnarray}

\section{Conclusions}

In this paper, we present a two-and-a-half dimensional symplectic space-charge
solver for a long charged particle beam bunch (in the beam frame).
Instead of solving the full three-dimensional Poisson equation,
this solver addresses the two-dimensional Poisson equation by exploiting
the slow longitudinal variation of the beam.
This approach significantly improves computational speed in
multi-particle tracking simulations where fast space-charge solver is
needed. Meanwhile, the longitudinal
momentum kick is included to account for longitudinal density
variation, in addition to the transverse momentum kicks.
As a result, a symplectic space-charge solver is achieved for
particle tracking.

For a transverse Gaussian density distribution, a semi-analytical
solution is obtained. By comparing this solution with that from
a fully three-dimensional Gaussian density distribution, we observe that
for a longitudinal-to-transverse aspect ratio (in the beam frame) greater
than $1000$, the zeroth-order two-and-a-half dimensional space-charge solver provides a good approximation to the full three-dimensional
solution.

For an arbitrary transverse density distribution, several efficient
numerical methods are presented under different boundary
conditions. These include an FFT-based integrated Green's
function method and two spectral methods. The computational cost
of the Green's function method scales as $O(N\log N)$, where $N$ is
the total number of computational grid points. The computational cost of the spectral method
for a rectangular perfectly conducting pipe also scales as $O(N\log N)$,
due to the use of the fast sine transform.
The computational cost of the spectral method for a round
conducting pipe is somewhat higher due to the use of
Bessel functions. However, given the rapid convergence of the
spectral method, only a small number of spectral modes are needed.

In addition to straight accelerator systems, the two-and-a-half
dimensional space-charge solver is extended to circular accelerators.
Using a transverse Gaussian density distribution, we show that
the correction to the electric potential due to the finite curvature
of a circular accelerator is small and is proportional to
the ratio of the beam size to the bending radius. For a typical
circular accelerator with a bending radius greater than $100$ meters,
this correction can be as small as $O(10^{-5})$.
Thus, the finite curvature
effect in the space-charge solver may be negligible in
large accelerators.

\section*{ACKNOWLEDGEMENTS}
We thank Yue Hao, Helena Alamprese and Chad Mitchell for their valuable discussions. This work was supported by the U.S. 
Department of Energy (DOE), Office of Science, Office of 
High Energy Physics, under Contract No. DE-AC02- 
05CH11231. Computational resources were provided by 
the National Energy Research Scientific Computing Center 
(NERSC), which is supported by the DOE Office of 
Science under the same contract number.
%This research was supported by the U.S. Department of Energy under Contract No. DE-AC02-05CH11231,
%and used computer resources at the National Energy Research
%Scientific Computing Center.

\section*{Appendix}
In the Appendix, we show Fortran and C++ implementations of the integrals in the space-charge solver
for a Gaussian density distribution and present the derivation of the first-order correction to the solution of the two-dimensional Poisson equation, accounting for the effect of
%effects of longitudinal variation in the three-dimensional Poisson equation and 
the finite curvature of the circular accelerator system, for a transverse Gaussian density distribution.

\subsection{Fortran implementation of integral in Eq. 25, Eq. 31, and Eq. 32}
\begin{verbatim}
!---------------------------------------------------------------------------------------------
!The following subroutine calculates three integrals used to determine the 
!3 space-charge kicks for a transverse Gaussian distribution in the paper. 
!Note: This implementation might be most efficient for a small aspect ratio beam (e.g. A<5). 
!
       !Inputs: delta - small region near 0 (e.g. 0.01), nint - # of grid points (odd number).
       !        xin, yin - x and y coordinates, sigx, sigy - x RMS size and y RMS size.
       !Outputs: pintex - integral for Ex in Eq.31; pintey - integral for Ey in Eq. 32;
       !         pintez - integral for Ez in Eq. 25. (including *2A)
       !Author:: Ji Qiang
!---------------------------------------------------------------------------------------------
       subroutine potInt(delta,nint,xin,yin,sigx,sigy,pintex,pintey,pintez)
       implicit none
       integer, intent(inout) :: nint
       real*8,intent(in) :: delta,xin,yin,sigx,sigy
       real*8,intent(out) :: pintex,pintey,pintez
       !xp is x/sigma_x, yp is y/sigma_y
       !asp = sigma_x/sigma_y
       real*8 :: asp,xp,yp,asp2,asp2m
       real*8 :: sum0,xmin,xmax,x,h,t2,f1,f2,fx,xp2,yp2,sum0ex,sum0ey
       real*8, dimension(nint+1) :: xvec,t2vec,t2sqrt
       real*8, dimension(nint+1) :: f2vec,f2yvec,f1vec,f1xvec,fxvec
       integer :: i
       real*8 :: exparg,sqrt2,f1x

       !enforce an odd grid points
       if(mod(nint,2).eq.0) nint = nint+1

       !xmin = 0.01d0
       xmin = delta
       xmax = 1.0d0
       h = (xmax-xmin)/(nint-1)

       xp = xin/sigx
       yp = yin/sigy
       asp = sigx/sigy

!----------------------------------------------------
! The following does integral Eq.25 without division of 4pi\epsilon0
!
       !trapzoidal rule approximation integral from 0 to delta
       x = xmin
       t2 = asp**2+(1-asp**2)*x**2
       sqrt2 = sqrt(t2)
       exparg = -(xp**2+yp**2/t2)/2
       f1 = x*exparg
       sum0 = x*f1/sqrt2/2

       !Ex
       f1 = x*exp(x**2*exparg)
       sum0ex = x*f1/sqrt2/2
       !Ey
       sum0ey = x*f1/(t2*sqrt2)/2

       !Simpson's rule for two end points
!-------------
       !Ez
       f1 = exp(x**2*exparg)
       f2 = x*sqrt2
       fx = (f1-1.0d0)/f2
       sum0 = sum0-fx*h

       f1x = x*f1
       f2 = sqrt2
       fx = f1x/f2
       sum0ex = sum0ex-fx*h

       f2 = t2*sqrt2
       fx = f1x/f2
       sum0ey = sum0ey-fx*h

       !Ez
       x = xmax
       t2 = asp**2+(1-asp**2)*x**2
       sqrt2 = sqrt(t2)
       exparg = -(xp**2+yp**2/t2)/2
       f1 = exp(x**2*exparg)
       f2 = x*sqrt2
       fx = (f1-1.0d0)/f2
       sum0 = sum0-fx*h
       !Ex
       f1x = x*f1
       f2 = sqrt2
       fx = f1x/f2
       sum0ex = sum0ex-fx*h
       !Ex
       f2 = t2*sqrt2
       fx = f1x/f2
       sum0ey = sum0ey-fx*h
!-------------

       do i = 1, nint
         xvec(i) = xmin + (i-1)*h
       enddo
       asp2 = asp**2
       asp2m = 1.0d0-asp**2
       do i = 1, nint
         t2vec(i) = asp2+asp2m*xvec(i)**2
       enddo
       do i = 1, nint
        t2sqrt(i) = sqrt(t2vec(i))
       enddo
       !for Ey error function
       do i = 1, nint
        f2yvec(i) = t2vec(i)*t2sqrt(i)
       enddo
       do i = 1, nint
        f2vec(i) = xvec(i)*t2sqrt(i)
       enddo

       xp2 = xp**2
       yp2 = yp**2
       do i = 1, nint
        f1vec(i) = exp(-xvec(i)**2*(xp2+yp2/t2vec(i))/2)
       enddo
       !for Ex error function
       do i = 1, nint
        f1xvec(i) = xvec(i)*f1vec(i)
       enddo

!----------------------------------------------------
       !for potential used in Ez
       do i = 1, nint
         fxvec(i) = (f1vec(i)-1.0d0)/f2vec(i)
       enddo
       do i = 1, nint, 2
         sum0 = sum0 + 2*fxvec(i)*h
       enddo
       do i = 2, nint, 2
         sum0 = sum0 + 4*fxvec(i)*h
       enddo
       pintez = 2*asp*sum0/3.0d0
!----------------------------------------------------
! The following does integral in Eq.31 for Ex
!
       !for Ex
       do i = 1, nint
         fxvec(i) = f1xvec(i)/t2sqrt(i)
       enddo
       do i = 1, nint, 2
         sum0ex = sum0ex + 2*fxvec(i)*h
       enddo
       do i = 2, nint, 2
         sum0ex = sum0ex + 4*fxvec(i)*h
       enddo
       pintex = 2*asp*xp*sum0ex/3.0d0/sigx

!----------------------------------------------------
! The following does integral in Eq.32 for Ey
!
       !for Ey
       do i = 1, nint
         fxvec(i) = f1xvec(i)/f2yvec(i)
       enddo
       do i = 1, nint, 2
         sum0ey = sum0ey + 2*fxvec(i)*h
       enddo
       do i = 2, nint, 2
         sum0ey = sum0ey + 4*fxvec(i)*h
       enddo
       pintey = 2*asp*yp*sum0ey/3.0d0/sigy

       end subroutine potInt
   
\end{verbatim}

\subsection{C++ implementation of integral in Eq. 25, Eq. 31, and Eq. 32}

\begin{verbatim}
/*---------------------------------------------------------------------------------------------
!The following subroutine calculates three integrals used to determine the 
!3 space-charge kicks for a transverse Gaussian distribution in the paper.
       !Inputs: delta - small region near 0 (e.g. 0.01), nint - # of grid points (odd number).
       !        xin, yin - x and y coordinates, sigx, sigy - x RMS size and y RMS size.
       !Outputs: pintex - integral for Ex in Eq.31; pintey - integral for Ey in Eq. 32;
       !         pintez - integral for Ez in Eq. 25. (including *2A)
       !Author:: Ji Qiang with the help of ChatGPT
!---------------------------------------------------------------------------------------------
*/
void potInt(double delta, int& nint, double xin, double yin, double sigx, double sigy,
            double& pintex, double& pintey, double& pintez)
{
    // enforce odd grid points
    if (nint % 2 == 0) nint += 1;

    double xmin = delta;
    double xmax = 1.0;
    double h = (xmax - xmin) / (nint - 1);

    double xp = xin / sigx;
    double yp = yin / sigy;
    double asp = sigx / sigy;

    double sum0, sum0ex, sum0ey;
    sum0 = sum0ex = sum0ey = 0.0;

    // Trapezoidal rule approximation integral from 0 to delta
    double x = xmin;
    double t2 = asp*asp + (1 - asp*asp)*x*x;
    double sqrt2 = std::sqrt(t2);
    double exparg = -(xp*xp + yp*yp / t2) / 2.0;
    double f1 = x*exparg;

    sum0 = x*f1/sqrt2/2.0;
    f1 = x*std::exp(x*x*exparg);
    sum0ex = x*f1 / sqrt2 / 2.0;
    sum0ey = x*f1 / (t2*sqrt2) / 2.0;

    // Simpson's rule for two end points
    // Ez
    f1 = std::exp(x*x*exparg);
    double f2 = x*sqrt2;
    double fx = (f1 - 1.0) / f2;
    sum0 = sum0 - fx*h;

    double f1x = x*f1;
    f2 = sqrt2;
    fx = f1x / f2;
    sum0ex = sum0ex - fx*h;

    f2 = t2*sqrt2;
    fx = f1x / f2;
    sum0ey = sum0ey - fx*h;

    // Ez at xmax
    x = xmax;
    t2 = asp*asp + (1 - asp*asp)*x*x;
    sqrt2 = std::sqrt(t2);
    exparg = -(xp*xp + yp*yp / t2) / 2.0;
    f1 = std::exp(x*x*exparg);
    f2 = x*sqrt2;
    fx = (f1 - 1.0) / f2;
    sum0 = sum0 - fx*h;
    f1x = x*f1;
    f2 = sqrt2;
    fx = f1x / f2;
    sum0ex = sum0ex - fx*h;
    f2 = t2*sqrt2;
    fx = f1x / f2;
    sum0ey = sum0ey - fx*h;

    // Arrays
    std::vector<double> xvec(nint), t2vec(nint), t2sqrt(nint);
    std::vector<double> f2vec(nint), f2yvec(nint), f1vec(nint), f1xvec(nint), fxvec(nint);

    for (int i = 0; i < nint; ++i)
        xvec[i] = xmin + i*h;
    double asp2 = asp*asp;
    double asp2m = 1.0 - asp*asp;
    for (int i = 0; i < nint; ++i)
        t2vec[i] = asp2 + asp2m * xvec[i]*xvec[i];
    for (int i = 0; i < nint; ++i)
        t2sqrt[i] = std::sqrt(t2vec[i]);
    for (int i = 0; i < nint; ++i)
        f2yvec[i] = t2vec[i]*t2sqrt[i];
    for (int i = 0; i < nint; ++i)
        f2vec[i] = xvec[i]*t2sqrt[i];

    double xp2 = xp*xp;
    double yp2 = yp*yp;
    for (int i = 0; i < nint; ++i)
        f1vec[i] = std::exp(-xvec[i]*xvec[i]*(xp2 + yp2 / t2vec[i]) / 2.0);
    for (int i = 0; i < nint; ++i)
        f1xvec[i] = xvec[i]*f1vec[i];

    // Ez
    for (int i = 0; i < nint; ++i)
        fxvec[i] = (f1vec[i] - 1.0) / f2vec[i];
    for (int i = 0; i < nint; i += 2)
        sum0 += 2 * fxvec[i] * h;
    for (int i = 1; i < nint; i += 2)
        sum0 += 4 * fxvec[i] * h;
    pintez = 2 * asp * sum0 / 3.0;

    // Ex
    for (int i = 0; i < nint; ++i)
        fxvec[i] = f1xvec[i] / t2sqrt[i];
    for (int i = 0; i < nint; i += 2)
        sum0ex += 2 * fxvec[i] * h;
    for (int i = 1; i < nint; i += 2)
        sum0ex += 4 * fxvec[i] * h;
    pintex = 2 * asp * xp * sum0ex / 3.0 / sigx;

    // Ey
    for (int i = 0; i < nint; ++i)
        fxvec[i] = f1xvec[i] / f2yvec[i];
    for (int i = 0; i < nint; i += 2)
        sum0ey += 2 * fxvec[i] * h;
    for (int i = 1; i < nint; i += 2)
        sum0ey += 4 * fxvec[i] * h;
    pintey = 2 * asp * yp * sum0ey / 3.0 / sigy;
}

\end{verbatim}

\subsection{Fortran implementation of 3D field integrals}
\begin{verbatim}
!---------------------------------------------------------------------------------------------
!The following subroutine calculates three integrals used to calculate the 
!3 space-charge kicks for a 3D Gaussian distribution in the paper.
!
       !Inputs: nint - # of grid points (odd number),
       !        xin, yin, zin - x, y and z coordinates, sigx, sigy, sigz -
       !        x RMS size, y RMS size, and z RMS size, gamma - relativistic factor.
       !Outputs: pintex - integral for Ex in Eq.45; pintey - integral for Ey in Eq. 46;
       !         pintez - integral for Ez in Eq. 47.
       !Author:: Ji Qiang
!---------------------------------------------------------------------------------------------
       subroutine efldgauss(nint,xin,yin,zin,sigx,sigy,sigz,gamma,pintex,pintey,pintez)
       implicit none
       integer, intent(inout) :: nint
       real*8,intent(in) :: xin,yin,zin,sigx,sigy,sigz,gamma
       real*8,intent(out) :: pintex,pintey,pintez
       !xp is x/sigma_x, yp is y/sigma_y, zp is z/sigma_z
       !asp = sigma_x/sigma_y
       !basp = sigma_x/(sigma_z*gamma)
       real*8 :: asp,xp,yp,asp2,asp2m,basp,zp
       real*8 :: xmin,xmax,x,h,t2,bt2,f1,f2,fx,xp2,yp2,zp2,sum0ex,sum0ey,sum0ez
       real*8, dimension(nint+1) :: xvec,t2vec,t2sqrt,bt2vec,bt2sqrt
       real*8, dimension(nint+1) :: f2xvec,f2yvec,f2zvec,f1vec,fxvec
       integer :: i
       real*8 :: exparg,sqrt2,sqrbt2,f1x

       !enforce an odd grid points
       if(mod(nint,2).eq.0) nint = nint+1

       xmin = 0.0d0
       xmax = 1.0d0
       h = (xmax-xmin)/(nint-1)
       xp = xin/sigx
       yp = yin/sigy
       zp = zin/sigz
       asp = sigx/sigy
       basp = sigx/(sigz*gamma)

       x = xmax
       t2 = asp**2+(1-asp**2)*x**2
       sqrt2 = sqrt(t2)
       bt2 = basp**2+(1-basp**2)*x**2
       sqrbt2 = sqrt(bt2)

       sum0ex = 0.0d0
       sum0ey = 0.0d0
       sum0ez = 0.0d0

       !end point correction
       !Ex
       exparg = -(xp**2+yp**2/t2+zp**2/bt2)/2
       f1 = x**2*exp(x**2*exparg)
       f2 = sqrt2*sqrbt2
       fx = f1/f2
       sum0ex = sum0ex-fx*h

       !Ey
       f2 = t2*sqrt2*sqrbt2
       fx = f1/f2
       sum0ey = sum0ey-fx*h

       !Ez
       f2 = bt2*sqrt2*sqrbt2
       fx = f1/f2
       sum0ez = sum0ez-fx*h
!-------------

       do i = 1, nint
         xvec(i) = xmin + (i-1)*h
       enddo
       do i = 1, nint
         t2vec(i) = asp**2+(1.0d0-asp**2)*xvec(i)**2
       enddo
       do i = 1, nint
        t2sqrt(i) = sqrt(t2vec(i))
       enddo
       do i = 1, nint
         bt2vec(i) = basp**2+(1.0d0-basp**2)*xvec(i)**2
       enddo
       do i = 1, nint
        bt2sqrt(i) = sqrt(bt2vec(i))
       enddo

       do i = 1, nint
        f2xvec(i) = t2sqrt(i)*bt2sqrt(i)
       enddo
       do i = 1, nint
        f2yvec(i) = t2vec(i)*t2sqrt(i)*bt2sqrt(i)
       enddo
       do i = 1, nint
        f2zvec(i) = bt2vec(i)*t2sqrt(i)*bt2sqrt(i)
       enddo

       xp2 = xp**2
       yp2 = yp**2
       zp2 = zp**2
       do i = 1, nint
        f1vec(i) = xvec(i)**2*exp(-xvec(i)**2*(xp2+yp2/t2vec(i)+zp2/bt2vec(i))/2)
       enddo

!----------------------------------------------------
       !for Ex
       do i = 1, nint
         fxvec(i) = f1vec(i)/f2xvec(i)
       enddo
       do i = 1, nint, 2
         sum0ex = sum0ex + 2*fxvec(i)*h
       enddo
       do i = 2, nint, 2
         sum0ex = sum0ex + 4*fxvec(i)*h
       enddo
       pintex = sum0ex/3.0d0

       !for Ey
       do i = 1, nint
         fxvec(i) = f1vec(i)/f2yvec(i)
       enddo
       do i = 1, nint, 2
         sum0ey = sum0ey + 2*fxvec(i)*h
       enddo
       do i = 2, nint, 2
         sum0ey = sum0ey + 4*fxvec(i)*h
       enddo
       pintey = sum0ey/3.0d0

       !for Ez
       do i = 1, nint
         fxvec(i) = f1vec(i)/f2zvec(i)
       enddo
       do i = 1, nint, 2
         sum0ez = sum0ez + 2*fxvec(i)*h
       enddo
       do i = 2, nint, 2
         sum0ez = sum0ez + 4*fxvec(i)*h
       enddo
       pintez = sum0ez/3.0d0

!----------------------------------------------------
       end subroutine efldgauss

\end{verbatim}
\subsection{C++ implementation of 3D field integrals}
\begin{verbatim}
/*----------------------------------------------------------------------------------------------
!The following subroutine calculates three integrals used to calculate the 
!3 space-charge kicks for a 3D Gaussian distribution in the paper.
!
       !Inputs: nint - # of grid points (odd number),
       !        xin, yin, zin - x, y and z coordinates, sigx, sigy, sigz -
       !        x RMS size, y RMS size, and z RMS size, gamma - relativistic factor.
       !Outputs: pintex - integral for Ex in Eq.45; pintey - integral for Ey in Eq. 46;
       !         pintez - integral for Ez in Eq. 47.
       !Author:: Ji Qiang with help of ChatGPT
!---------------------------------------------------------------------------------------------
*/
void efldgauss(int &nint, double xin, double yin, double zin,
               double sigx, double sigy, double sigz, double gamma,
               double &pintex, double &pintey, double &pintez)
{
    // enforce an odd grid points
    if (nint % 2 == 0) nint += 1;

    double xmin = 0.0, xmax = 1.0;
    double h = (xmax - xmin) / (nint - 1);
    double xp = xin / sigx;
    double yp = yin / sigy;
    double zp = zin / sigz;
    double asp = sigx / sigy;
    double basp = sigx / (sigz * gamma);

    std::vector<double> xvec(nint), t2vec(nint), t2sqrt(nint), bt2vec(nint), bt2sqrt(nint);
    std::vector<double> f2xvec(nint), f2yvec(nint), f2zvec(nint), f1vec(nint), fxvec(nint);

    for (int i = 0; i < nint; ++i)
        xvec[i] = xmin + i * h;

    for (int i = 0; i < nint; ++i)
        t2vec[i] = asp * asp + (1.0 - asp * asp) * xvec[i] * xvec[i];

    for (int i = 0; i < nint; ++i)
        t2sqrt[i] = std::sqrt(t2vec[i]);

    for (int i = 0; i < nint; ++i)
        bt2vec[i] = basp * basp + (1.0 - basp * basp) * xvec[i] * xvec[i];

    for (int i = 0; i < nint; ++i)
        bt2sqrt[i] = std::sqrt(bt2vec[i]);

    for (int i = 0; i < nint; ++i)
        f2xvec[i] = t2sqrt[i] * bt2sqrt[i];

    for (int i = 0; i < nint; ++i)
        f2yvec[i] = t2vec[i] * t2sqrt[i] * bt2sqrt[i];

    for (int i = 0; i < nint; ++i)
        f2zvec[i] = bt2vec[i] * t2sqrt[i] * bt2sqrt[i];

    double xp2 = xp * xp, yp2 = yp * yp, zp2 = zp * zp;
    for (int i = 0; i < nint; ++i)
        f1vec[i] = xvec[i] * xvec[i] * std::exp(-xvec[i] * xvec[i] * (xp2 + yp2 / t2vec[i] + zp2 / bt2vec[i
]) / 2.0);

    // Simpson's rule integration
    double sum0ex = 0.0, sum0ey = 0.0, sum0ez = 0.0;

    // Ex
    for (int i = 0; i < nint; ++i)
        fxvec[i] = f1vec[i] / f2xvec[i];
    for (int i = 0; i < nint; i += 2)
        sum0ex += 2.0 * fxvec[i] * h;
    for (int i = 1; i < nint; i += 2)
        sum0ex += 4.0 * fxvec[i] * h;
    //end point correction
    sum0ex -= fxvec[nint-1] * h;
    pintex = sum0ex / 3.0;

    // Ey
    for (int i = 0; i < nint; ++i)
        fxvec[i] = f1vec[i] / f2yvec[i];
    for (int i = 0; i < nint; i += 2)
        sum0ey += 2.0 * fxvec[i] * h;
    for (int i = 1; i < nint; i += 2)
        sum0ey += 4.0 * fxvec[i] * h;
    sum0ey -= fxvec[nint-1] * h;
    pintey = sum0ey / 3.0;

    // Ez
    for (int i = 0; i < nint; ++i)
        fxvec[i] = f1vec[i] / f2zvec[i];
    for (int i = 0; i < nint; i += 2)
        sum0ez += 2.0 * fxvec[i] * h;
    for (int i = 1; i < nint; i += 2)
        sum0ez += 4.0 * fxvec[i] * h;
    sum0ez -= fxvec[nint-1] * h;
    pintez = sum0ez / 3.0;
}

\end{verbatim}

\subsection{Derivation of the first-order curvature correction}
The first-order correction to the two-dimensional Poisson equation in a circular 
accelerator system can be written as:
\begin{eqnarray}
\frac{\partial^2 \phi_1}{\partial x^2} +
\frac{\partial^2 \phi_1}{\partial y^2} &  =  & - \frac{\rho(x,y,z)}{\epsilon_0}  
- h\frac{\partial \phi_0}{\partial x} 
\end{eqnarray}
%Following the same procedure as in the previous section, and 
Letting $\phi_1 = \phi_0 + \tilde{\phi}_1$, the above equation can be reduced to:
\begin{eqnarray}
 \frac{\partial^2 \phi_0}{\partial x^2} +
\frac{\partial^2 \phi_0}{\partial y^2} & = &- \frac{n(x,y)\lambda(z)}{\epsilon_0}  \\
\frac{\partial^2 \tilde{\phi}_1}{\partial x^2} +
\frac{\partial^2 \tilde{\phi}_1}{\partial y^2} &  =  & 
- h\frac{\partial \phi_0}{\partial x} 
\end{eqnarray}
The above equations can be solved by applying the Fourier transform, for 
a transverse bi-Gaussian density distribution, which yields:
\begin{eqnarray}
    \tilde{\Phi}_{1}(k_x,k_y) & = & h\frac{ik_x }{\epsilon_0 (k_x^2+k_y^2)^2}\exp(-\frac{\sigma_x^2 k_x^2}{2}-\frac{\sigma_y^2 k_y^2}{2})
\end{eqnarray}
Here, we have used
\begin{eqnarray}
    \tilde{\phi}_{1}(x,y) & = & \frac{1}{4\pi^2}\int_{-\infty}^{\infty}dk_x\int_{-\infty}^{\infty}dk_y \tilde{\Phi}_{1}(k_x,k_y) \exp(ik_x x+ik_y y)
\end{eqnarray}
and
\begin{eqnarray}
    {\Phi}_{0}(k_x,k_y) & = & h\frac{\lambda(z) }{\epsilon_0 (k_x^2+k_y^2)}\exp(-\frac{\sigma_x^2 k_x^2}{2}-\frac{\sigma_y^2 k_y^2}{2})
\end{eqnarray}
Taking the inverse Fourier transform and making use of the
following identities:
\begin{eqnarray}
\frac{1}{(k_x^2+k_y^2)^2} & = &\int_0^{\infty} t \exp(-t(k_x^2+k_y^2)) dt \\
   \int_{-\infty}^{\infty}\exp(-Ax^2+Bx) dx & = & \sqrt{\frac{\pi}{A}}\exp(\frac{B^2}{4A}) \\
   \int_{-\infty}^{\infty}x\exp(-Ax^2+Bx) dx & = & \frac{B}{2A}\sqrt{\frac{\pi}{A}}\exp(\frac{B^2}{4A})
\end{eqnarray}
we obtain
\begin{eqnarray}
    \tilde{\phi}_1(x,y,z) & = & - h\frac{\lambda(z) x}{8 \pi \epsilon_0}\int_0^{\infty} t dt
    \frac{\exp[-\frac{x^2}{2(\sigma_x^2+t)}-\frac{y^2}{2(\sigma_y^2+t)}]}{(\sigma_x^2+t)\sqrt{(\sigma_x^2+t)(\sigma_y^2+t) }    }
\end{eqnarray}
By introducing a linear $x$ term, the regularized solution~\ref{sol2} is obtained:
\begin{eqnarray}
    \tilde{\phi}_1(x,y,z) & = & - h\frac{\lambda(z) x}{8 \pi \epsilon_0}\int_0^{\infty} t dt
    \frac{\exp[-\frac{x^2}{2(\sigma_x^2+t)}-\frac{y^2}{2(\sigma_y^2+t)}] - 1}{(\sigma_x^2+t)\sqrt{(\sigma_x^2+t)(\sigma_y^2+t) }   }
\end{eqnarray}

\end{document}